\documentclass[preprint,12pt]{elsarticle}

\usepackage{calc}
\usepackage{color}
\usepackage{amsfonts}
\usepackage{stmaryrd}
\usepackage{float}
\usepackage{mathrsfs}
\usepackage{hyperref}
\usepackage{latexsym}
\usepackage{placeins}
\usepackage{ifthen}
\usepackage{amssymb}
\usepackage{amsmath}
\usepackage{framed}
\usepackage[cp1250]{inputenc}

\newtheorem{defi}{Definition}[section]
\newtheorem{tw}{Theorem}[section]

\newtheorem{lemat}{Lemma}[section]
\newtheorem{uwaga}{Remark}[section]

\newcommand{\mcal}{\mathcal}
\newcommand{\mbb}{\mathbb}
\newcommand{\mbf}{\mathbf}
\newcommand{\bydefi}{\colon\!\!=}%{\stackrel{\,\Delta}{=}}

\newcommand{\tinT}{{t\in\mbb{T}}}

\newcommand{\dowod}{\noindent{\bf Proof}:{\ }}
\newcommand{\koniec}{\newline  \mbox{}\hfill$\Box$}

\newcommand{\condE}[4]{\mbf{E}\left.\left[#1\right|\mcal{#2}^{#3}_{#4}\right]}
\newcommand{\barcondE}[4]{\bar{\mbf{E}}\left.\left[#1\right|\mcal{#2}^{#3}_{#4}\right]}
\newcommand{\dd}{\mathrm{d}}
\newcommand{\Proj}{\mathrm{Proj}}
\newcommand{\ATTENTION}[1]{}
\newcommand{\ATT}[1]{}

\newcommand{\wstawrysunek}[7]{
\begin{figure}\label{#2}
\begin{minipage}{\textwidth}
\caption{#3}
\begin{center}
\ifx\pdftexversion\undefined
\includegraphics[width=13cm]{#1}
\else
\includegraphics[trim = 100 150 100 150,height=7cm,angle=90]{#1}
\fi
\end{center}\ifthenelse{\equal{#7}{\empty}}{}{{\small #7}\\ }\ifthenelse{\equal{#5}{\empty}}{\small Source:\ #4}{{\small #6:\ #5}\\{\small Source:\ #4}}
\end{minipage}\end{figure}}

\newcommand{\wstawrysunekR}[7]{
\begin{figure}
\begin{minipage}{\textwidth}
\caption{#3}
\begin{center}
\ifx\pdftexversion\undefined
\includegraphics[width=13cm]{#1}\label{#2}
\else
\includegraphics[trim = 40 30 40 30,height=6cm]{#1}\label{#2}
\fi
\end{center}\ifthenelse{\equal{#7}{\empty}}{}{{\small #7}\\ }\ifthenelse{\equal{#5}{\empty}}{\small Source:\ #4}{{\small #6:\ #5}\\{\small Source:\ #4}}
\end{minipage}\end{figure}}

\newcommand{\wstawrysunekRhalf}[7]{
\begin{figure}
\begin{minipage}{\textwidth}
\caption{#3}
\begin{center}
\ifx\pdftexversion\undefined
\includegraphics[width=13cm]{#1}\label{#2}
\else
\includegraphics[trim = 0 265 0 20,clip=true,height=3cm]{#1}\label{#2}
\fi
\end{center}\ifthenelse{\equal{#7}{\empty}}{}{{\small #7}\\ }\ifthenelse{\equal{#5}{\empty}}{\small Source:\ #4}{{\small #6:\ #5}\\{\small Source:\ #4}}
\end{minipage}\end{figure}}

\newcommand{\wstawrysunekRb}[7]{
\begin{figure}
\begin{minipage}{\textwidth}
\caption{#3}
\begin{center}
\ifx\pdftexversion\undefined
\includegraphics[width=13cm]{#1}\label{#2}
\else
\includegraphics[trim = 0 0 0 0,height=6cm]{#1}\label{#2}
\fi
\end{center}\ifthenelse{\equal{#7}{\empty}}{}{{\small #7}\\ }\ifthenelse{\equal{#5}{\empty}}{\small Source:\ #4}{{\small #6:\ #5}\\{\small Source:\ #4}}
\end{minipage}\end{figure}}

\newcommand{\wstawrysunekRc}[7]{
\begin{figure}
\begin{minipage}{\textwidth}
\caption{#3}
\begin{center}
\ifx\pdftexversion\undefined
\includegraphics[width=13cm]{#1}\label{#2}
\else
\includegraphics[trim = 80 80 80 60,height=6cm]{#1}\label{#2}
\fi
\end{center}\ifthenelse{\equal{#7}{\empty}}{}{{\small #7}\\ }\ifthenelse{\equal{#5}{\empty}}{\small Source:\ #4}{{\small #6:\ #5}\\{\small Source:\ #4}}
\end{minipage}\end{figure}}

\newcommand{\wstawrysunekRs}[7]{
\begin{figure}
\begin{minipage}{\textwidth}
\caption{#3}
\begin{center}
\ifx\pdftexversion\undefined
\includegraphics[width=13cm]{#1}\label{#2}
\else
\includegraphics[trim = 0 70 0 60,height=6cm]{#1}\label{#2}
\fi
\end{center}\ifthenelse{\equal{#7}{\empty}}{}{{\small #7}\\ }\ifthenelse{\equal{#5}{\empty}}{\small Source:\ #4}{{\small #6:\ #5}\\{\small Source:\ #4}}
\end{minipage}\end{figure}}

\newcommand{\wstawzwordhor}[7]{
\begin{figure}
\begin{minipage}{\textwidth}
\caption{\small #3}
\begin{center}
\ifx\pdftexversion\undefined
\includegraphics[width=13cm]{#1}
\else
\includegraphics[trim = 120 590 120 65,height=5cm]{#1}
\fi
\end{center}\ifthenelse{\equal{#7}{\empty}}{}{{\small #7}\\ }\ifthenelse{\equal{#5}{\empty}}{\small Źródło:\ #4}{{\small #6:\ #5}\\{\small Source:\ #4}}\label{#2}
\end{minipage}\end{figure}}

\newcommand{\rysnapolstrhor}[7]{
\begin{minipage}{0.5\linewidth}
\caption{#3}
\begin{center}
\ifx\pdftexversion\undefined
\includegraphics[width=0.4\textwidth]{#1}\label{#2}
\else
\includegraphics[trim = 50 220 50 230,width=0.95\textwidth]{#1}\label{#2}
\fi
\end{center}\ifthenelse{\equal{#7}{\empty}}{}{{\small #7}\\ }\ifthenelse{\equal{#5}{\empty}}{\small Source:\ #4}{{\small #6:\ #5}\\{\small Source:\ #4}}
\end{minipage}
}

\newcommand{\wstawtabele}[7]{
\begin{table}
\small \caption{#3}
\begin{center}
\input{#1}\label{#2}
\end{center}
\ifthenelse{\equal{#7}{\empty}}{}{{\small #7}\\
}\ifthenelse{\equal{#5}{\empty}}{\small Source:\ #4}{{\small #6:\
#5}\\{\small Source:\ #4}}
\end{table}
}

\newcommand{\sectionmoj}[1]{%
  \ifhmode\par\fi
  \removelastskip
  \vskip 3ex\goodbreak
  \refstepcounter{section}%
  \noindent
  \leavevmode
  \begingroup
  \bfseries
  \thesection.\
  #1.\quad
  \endgroup
  \addcontentsline{toc}{section}{%
    \protect\numberline{\thesection}%
    #1}%
}

\renewcommand{\subsection}[1]{%
  \ifhmode\par\fi
  \removelastskip
  \vskip 3ex\goodbreak
  \refstepcounter{subsection}%
  \noindent
  \leavevmode
  \begingroup
  \bfseries
  \thesubsection.\
  #1.\quad
  \endgroup
  \addcontentsline{toc}{subsection}{%
    \protect\numberline{\thesubsection}%
    #1}%
}

\definecolor{czrw}{rgb}{1,0,0}
\definecolor{darkgray}{gray}{.75}

\begin{document}
\begin{frontmatter}
\author{Grzegorz Ha\l aj}\ead{grzegorz.halaj@n-s.pl}
\address{Bank Pekao SA, ALM Office, Warsaw\\Postdoc fellow at the Fields Institute, Toronto}
\date{April 14, 2008}
\title{Density quantization method in the optimal portfolio choice with partial observation of stochastic volatility\tnoteref{t1}}
\tnotetext[t1]{DISCLAIMER: The paper may contain personal views or opinions of the author that are not necessarily
those of the Bank Pekao SA}

\bibliographystyle{plainnat}
\begin{abstract}Computational aspects of the optimal
consumption and investment with the partially observed stochastic volatility of the asset prices are considered. The new
quantization approach to filtering -- {\it density
quantization} -- is introduced which reduces the original infinite dimensional state
space of the problem to the finite quantization set. The density
quantization is embedded into the numerical algorithm to solve the
dynamic programming equation related to the portfolio optimization.
\end{abstract}
\end{frontmatter}
\section{Introduction}\label{sec:intro}
The optimal investment portfolio and consumption path choice have been studied extensively for about 5 decades but still poses computational difficulties. The standard Merton-like investment problem evolved into many research fields and one of them are the \textit{partial observation} models studied by \citet{lakner:1995,pham:2001,Sass,Stettner,Baghery,Halaj,brendle:mean_rev_unob}. Those models are imbedded into the control theory framework with partial information \citep{bensoussan}. In the case of portfolio choice in the partial observation, the investor does not observe all the factors defining the dynamics of the market. Hence, the optimal strategies are not only functions of the state variables like the wealth of the investor,
but depend on investor's beliefs about the true value of asset prices and economic
factors \citep{Stettner}. Those beliefs are represented by the probability
distribution of values of the factors given available information on
the market. This makes the value function of the investor to be defined on
the infinite dimensional state space. Following the quantization approach we introduced the numerical
method to solve the related dynamic programming equation.

The partial observation shows up in the portfolio models in two forms. The
first one is related to trend parameters of assets prices, the
second one to volatility. Models with an unobserved trend are
usually easy to reformulate in a full observation setting (eg. by
the change of measure approach the drift parameter of diffusion
price processes vanish). The portfolio problems with unknown
volatility are, however, more realistic and we assumed in our model
that the volatility of stocks is unobserved, also driven by some
unobserved economic factors. This is a standard assumption in
advanced models of volatility \citep{Barndorff,Elliot}.

We consider the numerical approximation of the solution to the following consumption and investment problem. For clarity of the presentation of the main idea we postpone the rigorous assumptions to the section \ref{sec:assumpt}. Let us denote by $\mbb{T}$ the set
$\{0,1,\dots,T\}\subset\mbb{N}\cap\{0\}$. We consider the probability
space $(\Omega,\mcal{F},\mbf{P})$ and the discrete filtration
$\mbb{F}\bydefi\{\mcal{F}_t\}_{\tinT}$. On $\mcal{F}$ we define $N$
sequences of the IID random variables $\{\epsilon_i(t)\}_\tinT$,
$i\in\{1,\dots,N\}$, each of which has a positive density $\phi$. By
$\{r(t)\}_\tinT$ we denote a process of the interest rates in the
economy, constant within the periods $[t,\,t+1]$ and by $\{S_0(t)\}_\tinT$
we denote the bank account process given by the recurrence $S_0(0)=1$
and $\forall n>0\,S_0(t+1)=(1+r(t))S_0(t)$.

We assume that the process of the economic factors 
measuring the economic situation is denoted by $\{Y(t)\}_\tinT$ and
given by the following equation:
\begin{equation}\label{eq:unobs_fact_Y}
Y(0)=y_0\qquad
Y(t+1)=Y(t)+\alpha\left(\bar{y}-Y(t)\right)+\sigma_Y\xi(t+1),
\end{equation}
where $(\xi(t))_{t\in\mbb{T}}$ is the IID sequence of $K$-dimensional
random variables, each with positive density $\psi$. $\sigma_Y$ is the
full-rank $K\times K$ matrix. The tradable assets on the market have the
following dynamics for $i\in\{1,\dots,N\}$:
\begin{equation}\label{eq:assets_S_i}
S_i(t+1)=S_i(t)\exp\left(\mu(Y(t+1))+\sigma(Y(t+1))\epsilon_i(t+1)\right).
\end{equation}
There are two types of shocks on the market: the idiosyncratic
$\epsilon$ which are asset (eg. company) specific and the systemic $\xi$
which influence all the risky assets on the market.

The investor (and consumer at the same time) have preferences
characterized by utility functions
$u\colon\mbb{R}^K\times\mbb{R}_+\to\mbb{R}$ and
$u_T\colon\mbb{R}^K\times\mbb{R}_+\to\mbb{R}$.

The return from investment in the asset $S_i$ in $t$ for $[t,t+1]$ is
denoted by $R_i(t+1)={S_i(t+1)\over S_i(t)}$. The logarithmic return is
denoted $R_i^{l}(t)=\ln(R_i(t))$. The wealth of the investor starting
with capital $x_0$ and \textit{investing}
$\pi\bydefi\{\pi(t)|t\in\mbb{T}\}$ and \textit{consuming}
$c\bydefi\{c(t)|t\in\mbb{T}\}$ is defined by:
\begin{equation}
X(t+1)=\sum_{i=1}^N\pi_i(t){S_i(t+1)\over
S_i(t)}+(X(t)-\sum_{i=1}^N\pi_i(t)-c(t))(1+r(t)),\quad X(0)=x_0.
\end{equation}
The investor observes only the prices $S_i$ of the securities so his
information is described by filtration
$\mbb{F}^{S}=\{\mcal{F}^S_t\}_{\tinT}$ where
$\mcal{F}^{S}_t\bydefi\sigma(\{S_i(s)|i\in\{1,\dots,N\},\,s\leq
t\})$. It is equivalent to observing
$\mbb{F}^{lR}\bydefi\sigma(\{R_i^l(s)|i\in\{1,\dots,N\},\,s\leq
t\})$ generated by log-returns from investment in $S$ which is
more convenient for our purpose. The investor
wants to optimize his utility from consumption and his terminal wealth.
She maximizes the functional $J\colon\mcal{A}(x)\to\mbb{R}$ given by:
\begin{equation}\label{eq:functionalJ}
J(c,\pi)=\mbf{E}\left\{\sum_{t=1}^{T-1}\delta^{t}u(c(t))+\delta^{T}u_T(Y(T),X^{c,\pi}(T))\right\},
\end{equation}
where $T$ is the time horizon and $\mcal{A}(x)$ is the set of the admissible controls $(c,\,\pi)$
\begin{eqnarray}
\lefteqn{\mcal{A}(x)=\Big\{((c(0),\dots,c(T-1)),(\pi(0),\dots,\pi(T-1))|}\nonumber\\
&&(c(t),\pi(t))-\mcal{F}^{lR}_t\mathrm{\!-\!measurable};\qquad 0\leq c(t)\leq X^{c,\pi}(t);\nonumber\\&&\sum_{i=1}^N\pi_i(t)\leq
X^{c,\pi}(t)-c(t)\Big\}.\nonumber
\end{eqnarray}

The solution maximizing $J$ is given by the special form of dynamic programming.
\begin{eqnarray}
V(t,x,\varrho)&=&\sup_{\mcal{A}(x)}\Bigg\{u(\bar{c})+\delta\mbf{E}V\Big(t+1,x(\hat{R}^l(t+1)),\varrho_{t+1}(t+1,\hat{R}^l(t+1))\Big)\Bigg\}\nonumber\label{eq:value_recursIntro}
\end{eqnarray}
The arguments $t$ and $x$ refer to usual time and current wealth state space variables (like in \citet{Bertsekas}). The variable $\rho$ measures the distribution of a random variable which is precisely defined in section \ref{sec:filtering} and corresponds to the process $Y$. Unlike in the full observation stochastic control in which case the value function at time $t$ depends on the realization of $Y(t)$, the partial observation allows the investor to assess only the most likely value of $Y(t)$ given the history of the asset prices. That is the intuition behind the dependence of $V$ on the whole distribution of $Y$ given $\mcal{F}^S_t$.

The lack of closed form solutions to consumption/investment problems
gives rise to research on numerical simulation methods to be applied
to models with partial observation. The numerical procedure has to be designed in a special way to account for the natural high complexity of the state space of stochastic control. The literature does not give
many examples of approximate solutions to the stochastic control
with filtering. \citet{Runggaldier} discretize transition
probabilities and \citet{Desai} proposed a particle filtering method
to discretize a dynamic programming equation. There are numerous
studies of numerical methods of filtering equations themselves. The
Zakai equations which describe densities of unobserved parameters
evolving in time as new information arrives to the market can be
solved by means of the so-called particle filter methods
\citep{DelMoral}. Filter equations in discrete time can be estimated
with the so-called {\it quantization} \citep{Pages0,Pages,Pages2}. In the related paper \citet{Corsi} also proposed the scheme to solve the dynamic programming equation of the partially observed control problem applying markovian quantization and approximated the given Markov Chain by the transition probability matrix. In this way, infinite dimensional random variables can be approximated by a
finite set of point in an optimal way. We modified the standard
quantization techniques to develop a different approach -- {\it density quantization} -- and we
applied it to our control problem. Basically, the main idea is to find the optimal quantization set directly in the set of the unnormalized densities of the filter. In other words, the density quantization set is the set of the appropriately selected unnormalized density functions. It proved to be successful in
recursively solving the dynamic programming equation.

The paper is organized as follows: First, we present the
mathematical model of the market and decision making. Second, we
show the influence partial information can exert on consumption of
an investor in a simplistic theoretical example. Third, we rewrite
the partially observed problem in a full observation setting,
applying the change of measure approach and we solve it using the
dynamic programming. Fourth, we define the density quantization and
we show how the investment/consumption problem can be solved
numerically. Finally, we illustrate the application of the developed numerical method in the example of the market model.
\subsection{Notation}
\begin{itemize}
\item $\mcal{M}^{n\times m}$ --- the set of matrices $M$ with $n$ rows and $m$ columns.
\item $B^n(\mcal{F})$ --- the set of Borel functions
$f\colon\Omega\times\mbb{R}^K\to\mbb{R}$,
$f(\omega,\cdot)\in L^n(\mbb{R}^K)$.
\item $\phi_{\mu,\sigma}(x)$ --- Gaussian density i.e. $$\phi_{\mu,\sigma}(x)={1\over(2\pi)^{N/2}(\det(\Sigma))^{1/2}}\exp(-{1\over 2}(x-\mu)^{T}\Sigma^{-1}(x-\mu)),$$ where $\mu$ is mean and $\Sigma\bydefi\sigma\sigma^{\top}$ is covariance matrix.
\item $\mathrm{vol}(A)$ --- volume of a set $A\in\mbb{R}^n$.
\item $\mathrm{diam}(A)=\sup_{(x,y)\in A\times A}(\sum_{1\leq i\leq n}|x_i-y_i|^2)^{1/2}$.
\item $A^c$ --- completion of set $A$.
\end{itemize}
\subsection{Main technical assumptions}\label{sec:assumpt}
To show convergence of solutions of approximate problem to the
original one we need 3 assumptions for $\phi$ and $\psi$.\newline
[A0] $\sigma(\cdot)$ is bounded from 0, i.e.
$\exists\sigma_{\mathrm{min}}>0$ such that $\forall y\in\mbb{R}^K$\
$\sigma(y)>\sigma_{\mathrm{min}}$.
\newline [A1] $u$ and $u_T$ satisfy
$\int_{\mbb{R}^N\times\mbb{R}^K}(\phi(x)+\psi(y))(u(\exp(x))+u_T(\exp(x),y))\dd
x\dd y<\infty$.
\newline
[A2] For every $y\in\mbb{R}^N$ and $a<1$,
$\int_{\mbb{R}^K}\phi({z-y\over a}){1\over\phi(y)}\dd z<L_{\Phi}$.

{\em [A0]} and {\em [A2]} are necessary to show convergence of the
numerical scheme. {\em [A1]} guarantees the integrability in the
dynamic programming equations.
\section{Filtering}\label{sec:filtering}
The most fundamental task to be done before applying the standard
dynamic programming toolkit to our consumption/investment selection is to
transform the problem into the complete observation one. We apply
changing of reference measure techniques similarly to
\citet{Krishnamurthy,Elliot,Stettner}. After that, in the section
\ref{sec:dynProg}, we transform the classical dynamic programming
equation to make it suitable for our portfolio problem.

For each $t\in\mbb{T}$ we define a random variable
$$\lambda_t={\phi\left(\sigma^{-1}(Y(t))[R^{(l)}(t)-\mu(Y(t))]\right)\over\det(\sigma(Y(t)))\phi(R^{(l)}(t))}{\psi\Big(\sigma_Y^{-1}[Y(t)-Y(t-1)-\alpha(\bar{y}-Y(t-1))]\Big)\over\det(\sigma_Y)\psi(Y(t))}.$$
Let us construct a process $\Lambda$ in the following way:
$\Lambda_t\bydefi\prod_{s=1}^t\lambda_s.$

The process $\Lambda$ defines a new measure -- the {\it reference
measure} $\bar{\mbf{P}}$. Let us define the filtration
$\mbb{F}^{lR,Y}\colon\sigma(\{R^{l}(s),Y(s)|i\in\{1,\dots,N\},\,s\leq
t\})$. For each $t\in\mbb{T}$, its restriction to
$\mcal{F}^{lR,Y}_t$ is defined:
$${\dd\mbf{P}\over\dd\bar{\mbf{P}}}\Big|_{\mcal{F}^{lR,Y}_t}=\Lambda_t.$$
To study properties of stochastic processes under $\bar{\mbf{P}}$ we
extensively use the generalized Bayes formula: for
$(\mcal{F},\mbf{P})$-measurable $Z$
\begin{eqnarray}
\barcondE{Z}{F}{Y}{t}={\condE{\Lambda^{-1}_tZ}{F}{Y}{t}\over\condE{\Lambda^{-1}_t}{F}{Y}{t}}\quad\mathrm{and}\quad\condE{Z}{F}{Y}{t}={\barcondE{\Lambda_tZ}{F}{Y}{t}\over\barcondE{\Lambda_t}{F}{Y}{t}}.\label{eq:bayes_fromula}
\end{eqnarray}

We assume that there exists a function-valued process
$\varrho\colon\Omega\times\mbb{T}\times\mbb{R}^K\to L^2(\mbb{R})$ satisfying for a $L^2$ Borel
function $f$ we have
\begin{eqnarray}
\barcondE{\Lambda_tf(Y)}{F}{lR}{t}=\int_{\mbb{R}^K}f(z)\varrho_t(z)\dd
z,\ t\in\mbb{T}.
\end{eqnarray}
For a given $t$ the object $\varrho_t$ is the conditional unnormalized density of $Y$. We look for the recursive equation for $\varrho$ which --
combining with the Bayes formula \ref{eq:bayes_fromula} -- would
give the tractable expression for filters used in the optimization. It
makes the numerical simulation of the optimal consumption/investment
strategies easier.

The main result of the section, relying on independence of $Y$ and
$R^{l}$ under $\bar{\mbf{P}}$ is given in the following theorem.
\begin{tw}\label{tw:recursive_varrho}
Let us denote
\begin{eqnarray}
\Phi(z,R^{l}(t))&\bydefi&{\phi\Big(\sigma^{-1}(z)[R^{(l)}(t)-\mu(z)]\Big)
\over\det(\sigma(z))}\nonumber\\
\Psi(z,Y(t-1))&\bydefi&{\psi\Big(\sigma_Y^{-1}[z-Y(t-1)-\alpha(\bar{y}-Y(t-1))]\Big)\over\det(\sigma_Y)}.\nonumber
\end{eqnarray}
The process $\varrho$ satisfies the recursion
\begin{eqnarray}
\varrho_t(z)={\Phi(z,R^{l}(t))\over\phi(R^{l}(t))}\int_{\mbb{R}^K}\Psi(z,y)\varrho_{t-1}(y)\dd
y.\label{eq:recursive_filter}
\end{eqnarray}
\end{tw}
\dowod (see appendix \ref{app:filter})

Note that
$$\condE{f(Y(t))}{F}{lR}{t}={\int_{\mbb{R}^K}f(z)\varrho_t(z)\dd
z\over\int_{\mbb{R}^K}\varrho_t(z)\dd z}.$$ This representation allows us to rewrite the original
problem to the full observation control. Since $\Lambda$ is a
martingale then -- similarly to \citet{Runggaldier} -- we obtained
\begin{eqnarray}
\lefteqn{J(c,\pi)=\bar{\mbf{E}}\Lambda_T\left\{\sum_{t=1}^{T}\delta^{t}u(c(t))+\delta^{T}u_T(Y(T),X^{c,\pi}(T))\right\}}\nonumber\\
&&=\sum_{t=1}^{T}\bar{\mbf{E}}\left[\Lambda_t\delta^{t}u(c(t))\right]+\bar{\mbf{E}}\left[\Lambda_T\delta^{T}u_T(Y(T),X^{c,\pi}(T))\right]\nonumber\\
&&=\sum_{t=1}^{T}\bar{\mbf{E}}\left[\barcondE{\Lambda_t\delta^{t}u(c(t))}{F}{lR}{t}\right]+\bar{\mbf{E}}\left[\barcondE{\Lambda_T\delta^{T}u_T(Y(T),X^{c,\pi}(T))}{F}{lR}{T}\right]\nonumber\\
&&=\bar{\mbf{E}}\left\{\sum_{t=1}^{T}\delta^{t}u(c(t))+\delta^{T}\int_{\mbb{R}^K}u_T(z,X^{c,\pi}(T))\varrho_T(z)\dd
z\right\}.\label{eq:fromPartToFull}
\end{eqnarray}
In this way all the processes used in the equation \ref{eq:fromPartToFull} are observed, i.e. adapted to the filtration generated by the observed process $R^l$.
\begin{uwaga}
There is an analogical way of solving the filters in the continuous time.
The density of the filter is described by the so-called Zakai
equation. For instance, \citet{Benes} showed the existence of the
solution to the Zakai equation in the theoretical case and
\citet{Carmona} used this equation in the financial setting.
\end{uwaga}
\section{Dynamic programming}\label{sec:dynProg} After transcribing the original problem to a Markov one we
can formulate a dynamic programming equation for the optimal
wealth of an investor. Unfortunately, this is an
infinite-dimensional problem since the value functional of the
optimization is a function on the space of the random densities.

Let $\mcal{A}_t(x)$ be the set of the admissible controls of the portfolio
starting at time $t$ from wealth $x$ (initial at $t$)
i.e.
$$\mcal{A}_t(x)\bydefi\left\{\left((c(t),c(t+1),\dots,c(T)),(\pi(t),\pi(t+1),\dots,\pi(T)\right)|(c,\pi)\in\mcal{A}(x)\right\}.$$
{\it The optimal value} $V$ at time $t\in\mbb{T}$ with the initial
wealth $x$ and the updated density $\varrho$ of the unobserved
process at $t$ is defined $V\colon\mbb{T}\times\mbb{R}\times
B^2(\mcal{F})\to\mbb{R}$
\begin{eqnarray}
V(t,x,\varrho)\bydefi\sup_{(c,\pi)\in\mcal{A}_t(x)}\bar{\mbf{E}}\left[\sum_{s=t}^{T-1}\delta^{s-t}u(c(s))+\delta^{T-t}\int_{\mbb{R}^K}u_T(z,X^{c,\pi}(T))\varrho_T(z)\dd
z\Big|\mcal{F}_t^{lR}\right],\nonumber
\end{eqnarray}
for $X^{c,\pi}(t)\equiv x$ and $\varrho_t\equiv\varrho$. The
previous section gives
\begin{eqnarray}
V(T,x,\varrho)=\bar{\mbf{E}}\left[\int_{\mbb{R}^K}u_T(z,X^{c,\pi}(T))\varrho_T(z)\dd
z\Big|\mcal{F}_T^{lR}\right]=\int_{\mbb{R}^K}u_T(z,x)\varrho(z)\dd
z.\nonumber
\end{eqnarray}
\begin{lemat}
The following backward recursion holds for $V$ and $t<T$:
\begin{eqnarray}
V(t,x,\varrho)&=&\sup_{\begin{subarray}
\{\{(\bar{c},\bar{\pi})\in\mbb{R}\times\mbb{R}^N|0\leq \bar{c}\leq
x,\\ \quad \sum_{1\leq i\leq N}\bar{\pi}_i\leq
x\}\end{subarray}}\Bigg\{u(\bar{c})+\delta\int_{\mbb{R}^N}V\Big(t+1,\sum_{i=1}^N\bar{\pi}_i\hat{R}_i+\nonumber\\
&&+(x-\sum_{i=1}^N\bar{\pi}_i-\bar{c})(1+r(t)),\varrho_{t+1}\Big)\prod_{i=1}^N\phi(\hat{R}_i)\dd
\hat{R}_1\cdot...\cdot\dd \hat{R}_N\Bigg\}\label{eq:value_recursive}
\end{eqnarray}
and
\begin{eqnarray}
\sup_{(c,\pi)\in\mcal{A}}J(c,\pi)=V(0,x_0,\psi).\label{eq:JvsV}
\end{eqnarray}
\end{lemat}
\dowod It is straightforward that the identity \ref{eq:JvsV} hold.

By definition, $\mcal{F}^{lR}_t$-measurability of $(c(t),\pi(t))$
and $\mcal{F}_t^{lR}\subset\mcal{F}_{t+1}^{lR}$
\begin{eqnarray}
\lefteqn{V(t,x,\varrho)=\sup_{(c,\pi)\in\mcal{A}_t(x)}\Big\{\bar{\mbf{E}}\Big[\sum_{s=t}^T\delta^{s-t}u(c(s))+\delta^{T-t}\int_{\mbb{R}^K}u_T(z,X^{c,\pi})\varrho_T(z)\dd z\Big|\mcal{F}_t^{lR}\Big]\Big\}}\nonumber\\
&&=\sup_{(c,\pi)\in\mcal{A}_t(x)}\Big\{u(c(t))+\delta\bar{\mbf{E}}\Big[\sum_{s=t+1}^T\delta^{s-(t+1)}u(c(s))+\nonumber\\
&&\qquad\qquad\qquad\qquad\qquad\qquad\qquad\delta^{T-(t+1)}\int_{\mbb{R}^K}u_T(z,X^{c,\pi})\varrho_T(z)\dd z\Big|\mcal{F}_t^{lR}\Big]\Big\}\nonumber\\
&&=\sup_{(c,\pi)\in\mcal{A}_t(x)}\Big\{u(c(t))+\delta\bar{\mbf{E}}\Big[\bar{\mbf{E}}\Big[\sum_{s=t+1}^T\delta^{s-(t+1)}u(c(s))+\nonumber\\
&&\qquad\qquad\qquad\qquad\qquad\qquad\qquad\delta^{T-(t+1)}\int_{\mbb{R}^K}u_T(z,X^{c,\pi})\varrho_T(z)\dd z\Big|\mcal{F}_{t+1}^{lR}\Big]\mcal{F}_t^{lR}\Big]\Big\}\nonumber\\
&&=\sup_{(c,\pi)\in\mcal{A}_t(x)}\Big\{u(c(t))+\delta\bar{\mbf{E}}\Big[V(t+1,\varkappa(x,c(t),\pi(t),R^l(t+1)),\varrho_{t+1})\Big|\mcal{F}_t^{lR}\Big]\Big\}\nonumber
\end{eqnarray}
Using definition of $\Lambda_t$, it is easy to check that
$R^l_i(t+1)$ has the density $\phi$ under the measure $\bar{P}$. Let us define
$$\varkappa(t,\hat{x},\hat{c},\hat{\pi},\hat{R})=\sum_{i=1}^N\hat{\pi}_ie^{\hat{R}_i}+(\hat{x}-\sum_{i=1}^{N}\hat{\pi}_i-\hat{c})(1+r(t)).$$
\begin{eqnarray}
\lefteqn{=\sup_{(c,\pi)\in\mcal{A}_t(x)}\Big\{u(c(t))+\delta\bar{\mbf{E}}\Big[V\Big(t+1,\varkappa(x,c(t),\pi(t),R^l(t+1)),}\nonumber\\
&&\qquad\qquad\qquad\qquad\qquad\qquad\qquad\qquad{\Phi(\cdot,R^{l}(t+1))\over\phi(R^{l}(t+1))}\int_{\mbb{R}^K}\Psi(\cdot,y)\varrho_{t-1}(y)\dd
y\Big)\Big|\mcal{F}_t^{lR}\Big]\Big\}\nonumber\\
&&\sup_{(\bar{c},\bar{\pi})\in A(x)}\Big\{u(\bar{c})+\delta\int_{\mbb{R}^N}V\Big(t+1,\varkappa(x,\bar{c},\bar{\pi},\hat{R}),\nonumber\\
&&\qquad\qquad\qquad\qquad{\Phi(\cdot,\hat{R})\over\phi(\hat{R})}\int_{\mbb{R}^K}\Psi(\cdot,y)\varrho_{t-1}(y)\dd
y\Big)\prod_{i=1}^N\phi(\hat{R}_i)\dd\hat{R}_1\cdot...\cdot\dd
\hat{R}_N\Big\}.\nonumber
\end{eqnarray}
\koniec
\subsection{Quantization}
The set $B^2(\mcal{F})$ is infinite dimensional. Hence, $V$ in the form of the equation
\ref{eq:value_recursive} cannot be computed numerically. It has to
be approximated by a sequence of the functionals $\hat{V}$ on the finite state space. It is a
very complex numerical problem. To this end, \citet{Runggaldier}
proposed the scheme for the numerical filtering based on the Hidden Markov
Model approach. They discretized the infinite-dimensional equation
of the transition probabilities on the finite state space and then could
solve the dynamic programming equation on the set of vectors in
$\mbb{R}^M$. The grid was chosen arbitrarily. The partition in
this grid method is straightforward but easily gets untractable as
the number of nodes in the partition of the codomain increases. With
only 5 grids the number of the approximate value functions, which has to
be computed amounts to $5^5=3125$. 20 grids give
104857600000000000000000000 variants!

A much more applicable extension of this approach is the {\it density
quantization} introduced to limit the number of the unnormalized
densities in the set of arguments of the value function and to
optimize the structure of the grid.

To reduce dimensionality of
our problem from the infinite to finite space we follow the idea
proposed by \citet{Stettner}. We discretize the codomain of the
densities $\varrho$. We take a partition of each coordinate of
$\mbb{R}^K$, i.e. for the coordinate $j$ we take the set of $m_{n}^{(j)}$
increasing real numbers
$\{z_{n}^{(j)}(1),\dots,z_{n}^{(j)}(m_{n}^{(j)})\}$. We approximate
random densities $\varrho$ by the functional on the grid
$$Z_n\colon=\prod_{j=1}^K\{z_{n}^{(j)}(1),\dots,z_{n}^{(j)}(m_{n}^{(j)})\}\subset\mbb{R}^K.$$ The cubes
$B_n(k_1,\dots,k_K)\bydefi\prod_{j=1}^K[z_n^{(j)}(k_j),z_n^{(j)}(k_j+1)]$
give the partition of the support of $\rho_t$. The sum of the sets $B_n$ is
denoted $$\mathfrak{B}_n=\bigcup_{\{(k_1,\dots,k_K)|1\leq
k_i<m_{n}^{(i)}\}}B_n(k_1,\dots,k_K).$$ We assume that
\begin{eqnarray}
&&\lim_{n\to+\infty}\mathrm{vol}(\mathfrak{B}_n)=+\infty,\nonumber\\
\forall 1\leq j\leq
K&&\lim_{n\to+\infty}\max_{i\in\{1,\dots,m^{(j)}_{n}-1\}}|z_n^{(j)}(i+1)-z_n^{(j)}(i))|=
0.\nonumber
\end{eqnarray}
Similarly, we define the partition of the codomain of $\rho_t$:
$G_n\bydefi\{g_n(1),\dots,g_n(m_n)\}$ such that $g_n(i+1)>g_n(i)$,
$\lim_{n\to\infty}g_n=\infty$ and $\max_i|g_n(i+1)-g_n(i)|\to 0$ as
$n$ tends to $+\infty$.

The idea of the quantization is to focus attention on the set of density functions and to project all random
densities on this set. Unlike in the grid method where the projections
are made inside every subset $B_n$ on the grid $G_n$, the number of
outcomes can be quite small. Intuitively, under mild conditions the
densities form a ``tight'' family of functions and many of them are
very close to each other (in sup norm). There is no
point in distinguishing between them since they lead to similar
value functions. The idea of quantization gave fruitful results in
finding transition probabilities for nonlinear filters considered by
\citet{Pages}. It happens to be useful in the dynamic programming as well.

The fundamental idea of the approach is to project random densities
on a set of functions that have only finitely many values and vanish
at all point sufficiently distant from 0. The set of functions will
be called {\it density quantization set} (or shortly -- {\it
quantization set}).
\begin{defi}[Density quantization set]\label{defi:densQuantSet}
Any set of functions $q\colon\mbb{R}^K\to G_n$, which are constant
on the sets $B_n(k_1,\dots,k_K)$, $k_i\in\{1,\dots,m_n^{(i)}\}$ and
are equal to 0 on $\mathfrak{B}_n^c$ is called density quantization
set and is denoted $\mcal{Q}_n$. The number of elements of the set
$\mcal{Q}_n$ -- $\#\mcal{Q}_n$ -- is denoted $N_n$ and tends to
$+\infty$ with $n\to+\infty$.
\end{defi}

The projection $\mathrm{Proj}_{Z_n}^{\mcal{Q}_n}$ of a random
density $\varrho$ on $\mcal{Q}_n$ is defined
\begin{eqnarray}
\lefteqn{\mathrm{Proj}_{Z_n}^{\mcal{Q}_n}[\varrho]\bydefi{\arg\!\min}_{\{q\in\mcal{Q}_n\}}\max_{\{(k_1,\dots,k_K)|1\leq
k_i\leq
m_n^{(i)}\}}|\varrho(\omega,z_n^{(1)}(k_1),\dots,z_n^{(K)}(k_K))-}\nonumber\\&&\qquad\qquad\qquad\qquad\qquad\qquad\qquad\qquad\qquad\qquad q(z_n^{(1)}(k_1),\dots,z_n^{(K)}(k_K))|.\nonumber
\end{eqnarray}

The proper behavior of density quantization is guaranteed by the
so-called Zador Theorem \citep{Pages}
\begin{tw}\label{tw:Zador}
Let X be $n$-dimensional random variable such that $X\in
L^{2+\delta}$, $\delta>0$, with density $f_n$. Let $Q_K\subset\mbb{R}^n$ and the projection
$\mathrm{Proj}_{Q_K}\colon\mbb{R}\to Q_K$ be defined as
$\mathrm{Proj}_{Q_K}[x]={\arg\min}_{q\in Q_K}\|q-x\|$. Then
$$D_n(Q_K)\colon=\min_{\{Q_K|\#Q_K=K\}}\mbf{E}\|X-\mathrm{Proj}_{Q_K}(X)\|^2\leq
K^{-{2\over n}}J_n\cdot\left(\int_{\mbb{R}^n}f_n(x)^{n\over n+2}\dd
x\right)^{n+2\over n},$$ where $J_n\simeq{n\over 2\pi e}$.
\end{tw}
The quantity $D_n(Q_n)$ is called the distortion of the quantization.

The straightforward application of the Zador Theorem gives very slow
convergence. The structure of the stock market model allows to show
that, in fact, the numerical scheme have much better properties if we
assume Lipschitz continuity of $\Phi(z,\cdot)/\phi(\cdot)$.

Let us discretize the space of values of $R$ and consider the set
$Q_n^{(R)}\colon=\{\bar{r}_1,\dots,\bar{r}_n|\bar{r}_i\in\mbb{R}^N\}$.
Let the function $\bar{\varrho}$ be defined as
$$\bar{\varrho}(z,\bar{R})\colon=\Phi(z,\bar{R})/\phi(\bar{R})\int_{\mbb{R}^K}\Psi(z,y)\varrho(y)\dd
y.$$ From now on, we can think about $G_n$ as of the set {\it
generated} implicitly by $Q_n^{(R)}$, i.e.
$$G_n\equiv\{\bar{\varrho}(z_n^{(j)}(l),\bar{r}_i)|j\in\{1,\dots,n\},l\in\{1,\dots,m^{(j)}_n\},i\in\{1,\dots,n\}\}.$$
The quantization set from definition \ref{defi:densQuantSet},
corresponding to $Q_n^{(R)}$ is denoted $\mcal{Q}_{Q_n^{(R)}}$. The
set of all the sets $Q_n$ is denoted $\mcal{S}_Q$. A separate
discretization is defined for the portfolio process $X$, but in the
standard way followed in the numerical solving of the dynamic programming
equations (in the full observation setting or even in the
deterministic setting). The projection set is
$Q^X_n=\{x_1,\dots,x_n\}$ and the set of the projection sets is
denoted $\mcal{S}^X$.

Let us define the sequence of functionals $\widehat{V}$ on
$\mbb{T}\times \mbb{R}\times B^1(\mcal{F})$ as
\begin{eqnarray}
\widehat{V}(T,x,\varrho)&=&\int_{\mbb{R}^N}u_T(z,x)\varrho(z)\dd z\nonumber\\
\widehat{V}(t,x,\varrho)&=&\sup_{c,\pi\in\mcal{A}(t)}\big\{u(c)+\delta\bar{\mbb{E}}\widehat{V}(t+1,\mathrm{Proj}_{Q_X}[x^{c,\pi}(R)],\mathrm{Proj}^{\mathcal{Q}}[\varrho(R)])\big\}\label{eq:approxDynProg}
\end{eqnarray}
Note that $$V(T,x,\varrho)=\widehat{V}(T,x,\varrho)$$ for
$x\in\mcal{Q}^X$ and $\varrho\in\mcal{Q}$. In the following part of
the subsection we showed that $\widehat{V}$ approximates $V$.

For brevity, if it does not lead to confusion, let us denote
$\widehat{x}\bydefi\Proj_{\mcal{Q}_X}[x]$ and
$\widehat{\varrho}\bydefi\Proj_{\mcal{Q}}[\varrho]$.

\begin{lemat}\label{lemat:lipschVT}
If $u_T(\cdot,z)$ is $L_2$-Lipschitz continuous and bounded by $L_1$
then
\begin{eqnarray}
|V(T,x_1,\varrho_1)-\widehat{V}(T,x_2,\varrho_2)|^2\leq
2L^2_1\int_{\mbb{R}^K}|\varrho_1(z)-\widehat{\varrho}_2(z)|\dd
z+2L^2_2|x_1-\widehat{x}_2|.\nonumber
\end{eqnarray}
\end{lemat}
\dowod
\begin{eqnarray}
\lefteqn{|V(T,x_1,\varrho_1)-\widehat{V}(T,x_2,\varrho_2)|^2=\left|\int_{\mbb{R}^K}u(x_1,z)\varrho_1(z)\dd z-\int_{\mbb{R}^K}u(\widehat{x}_2,z)\widehat{\varrho}_2(z)\dd z\right|^2<}\nonumber\\
&&\int_{\mbb{R}^K}\left|u(x_1,z)\varrho_1(z)-u(x_1,z)\widehat{\varrho}_2(z)+u(x_1,z)\widehat{\varrho}_2(z)-u(\widehat{x}_2,z)\widehat{\varrho}_2(z)\right|^2\dd
z\leq\nonumber\\
&&2L_1^2\int_{\mbb{R}^K}|\varrho_1(z)-\widehat{\varrho_2}(z)|^2\dd
z+2L_2^2|x_1-\widehat{x}_2|^2.
\end{eqnarray}\koniec
\begin{defi}
The value function $v$ is $(\epsilon,c_0,c_1)$-Lipschitz continuous if
and only if there exist constants $c_0$ and $c_1$ such that for each
pair $(x_1,x_2)$ and $(\varrho_1,\varrho_2)$
$$|v(x_1,\varrho_1)-\widehat{v}(x_2,\varrho_2)|^2<\epsilon+c_0|x_1-x_2|^2+c_1\int_{\mbb{R}^K}|\varrho_1(z)-\varrho_2(z)|^2\dd
z.$$
\end{defi}
\begin{lemat}\label{tw:main_quantiz_result}
Let $\bar{\varrho}(z,\cdot)$ be Lipschitz continuous and is
$p$-decaying for $p>N$ i.e. $$\exists
a_z>0\quad\bar{\varrho}(z,x)<a_z|x|^{-p}.$$ Let us assume that
$\mathrm{vol}(B(\vec{k}))\leq \left({M_0\over n}\right)^K$. Let us
denote the distribution of $x^{(\bar{c},\bar{\pi})}_1(R)$ by
$f^{\bar{\pi}}_{X_1}\in L^{1\over 3}$ and
$x^{(\bar{c},\bar{\pi})}_2(R)$ by $f^{\bar{\pi}}_{X_2}\in L^{1\over
3}$. Then, for $(\epsilon,L_x,L_{\varrho})$-Lipschitz continuous
function $V$
\begin{eqnarray}
\lefteqn{\mathit{err}_n\colon=
\min_{\substack{\{Q^X\in\mcal{S}^X|\#Q^X=n\} \\
\{Q\in\mcal{S}_Q|\#Q=n\}}}\bar{\mbf{E}}\big|V(t,x^{(\bar{c},\bar{\pi})}_1(R),\varrho_1(R))-}\nonumber\\
&&\qquad\qquad\qquad\qquad\widehat{V}(t,\mathrm{Proj}_{Q_X}[x^{(\bar{c},\bar{\pi})}_2(R)],\mathrm{Proj}_{Z_n}^{\mcal{Q}_Q}[\varrho_2(R)])\big|^2\leq\nonumber\\
&&2L_x(1+r(t))^2|x_1-x_2|^2+2L_{\varrho}L_{\Phi}^2L_{\Psi}^2\int_{\mbb{R}^K}|\varrho_1(z)-\varrho_2(z)|^2\dd
z+\nonumber\\&&L_xn^{-2}J_1\cdot\left(\int_{\mbb{R}}f_X(x)^{1\over
3}\dd x\right)^3+2L_{\varrho}L_R^2\sqrt{K}\left({M_0\over
n}\right)^{K+1}+\nonumber\\&& \quad 4L_{\varrho}M_0^Kn^{-{2\over
N}}J_N\left(\int_{\mbb{R}^N}f^{max}_{z_k}(x)^{N\over N+2}\dd
x\right)^{N+2\over N}+2L_{\varrho}a_zv_N{M_0^{N-p}\over
p-1}+\epsilon,\qquad\qquad\nonumber
\end{eqnarray}
where
\begin{equation}\nonumber
v_N=\left\{\begin{matrix}{(2\pi)^{{N-2\over 2}+1}\over
(N-p)(N-2)!!},\hfill& \mathrm{if}\ N=2k\hfill\cr {2^{{N-3\over
2}+1}\pi^{{N-1\over 2}}\over (N-p)(N-2)!!},\hfill&\mathrm{if}\
N=2k+1\hfill\cr\end{matrix}\right.
\end{equation}
and $f^{max}_{z_k}$ is density of $\varrho(R)(z_k)$ for $k$ such
that
$$k={\arg\max}_{i}\left\{\left(\int_{\mbb{R}^N}f^{max}_{z_i}(x)^{N\over
N+2}\dd x\right)^{N+2\over N}\right\}.$$
\end{lemat}
\dowod By Lipschitz continuity of V and the the inequality $|a-b|^2\leq
2a^2+2b^2$
\begin{eqnarray}
\lefteqn{\Delta V\bydefi|V(t,x^{(\bar{c},\bar{\pi})}_1(R),\varrho_1(R))-\widehat{V}(t,\mathrm{Proj}_{Q_X}[x^{(\bar{c},\bar{\pi})}_2(R)],\mathrm{Proj}_{Z_n}^{\mcal{Q}_Q}[\varrho_2(R)])|^2\leq}\nonumber\\
&&\epsilon+2L_x|x_1^{(\bar{c},\bar{\pi})}(R)-\mathrm{Proj}_{Q}[x_2^{(\bar{c},\bar{\pi})}(R)]|^2+2L_{\varrho}\int_{\mbb{R}^K}|\varrho(R)(z)-\mathrm{Proj}_{Z_n}^{\mcal{Q}_Q}[\varrho(R)](z)|^2\dd z\leq\nonumber\\
&&4L_x|x_1^{(\bar{c},\bar{\pi})}(R)-x_2^{(\bar{c},\bar{\pi})}(R)|^2+4L_x|x_2^{(\bar{c},\bar{\pi})}(R)-\mathrm{Proj}_{Q_X}[x_2^{(\bar{c},\bar{\pi})}(R)]|^2+\nonumber\\
&&4L_{\varrho}\int_{\mbb{R}^K}\left|\varrho_1(R)(z)-\varrho_2(R)(z)\right|^2\dd
z+4L_{\varrho}\sum_{\vec{k}\in\hat{K}_n}\int_{B(\vec{k})}|\varrho_2(R)(z)-g_k|^2\dd z+\nonumber\\
&&4L_{\varrho}\int_{\mathfrak{B}^c_n}\varrho(R)^2(z)\dd z\nonumber
\end{eqnarray}
for a set of $g_k$s, $g_k\in G_n$.

Firstly, let us make an easy observation that
\begin{eqnarray}
|x_1^{(\bar{c},\bar{\pi})}(R)-x_2^{(\bar{c},\bar{\pi})}(R)|^2=(1+r(t))|x_1-x_2|^2.\nonumber
\end{eqnarray}
Secondly, we can use the quantization on $Q^X$ to estimate the
approximation of the random variable $x_2^{(\bar{c},\bar{\pi})}(R)$
with $\mathrm{Proj}_{Q_X}[x_2^{(\bar{c},\bar{\pi})}(R)]$. Since
$x^{(\bar{c},\bar{\pi})}(R)$ is the sum of the random variables of the
form $p_i\exp(R)$, $R\sim\phi(\cdot)$, then from the Zador theorem
(see theorem \ref{tw:Zador}) we get
\begin{eqnarray}
\lefteqn{\min_{\{Q^X\in\mcal{S}^X|\#Q^X=n\}}\bar{\mbf{E}}|x_2^{(\bar{c},\bar{\pi})}(R)-\mathrm{Proj}_{Q_X}[x_2^{(\bar{c},\bar{\pi})}(R)]|^2\leq}\nonumber\\
&&\qquad\qquad\max_{(p_1,\dots,p_N)\in[0,x_2]^N}n^{-2}J_1\left(\int_{\mbb{R}}\left(f^{p_1,...,p_N}_{X_2}(x)\right)^{1\over
3}\dd x\right)^3,\label{eq:zadorInAction}
\end{eqnarray}
where $f_{X_2}^{p_1,...,p_N}(\cdot)\bydefi
(\phi^{(e)}_{p_1}\ast...\ast\phi^{(e)}_{p_N})(\cdot)$ is the
convolution of densities of the random variables $p_i\exp(R)$. The
maximum always exists and we skipped the justification to the remark
\ref{rem:minimumExists}. We denoted the corresponding
$f^{p_1,...,p_N}_{X_2}$ by $f_X$.

Thirdly, by the assumption [A2]
\begin{eqnarray}
\lefteqn{\int_{\mbb{R}^K}\left|\varrho_1(R)(z)-\varrho_2(R)(z)\right|^2\dd z=}\nonumber\\
&&\int_{\mbb{R}^K}\left[{\Phi(z,R^{(l)})\over\phi(R^{(l)})}\int_{\mbb{R}^K}\Psi(z,y)\varrho_1(y)\dd
y-{\Phi(z,R^{(l)})\over\phi(R^{(l)})}\int_{\mbb{R}^K}\Psi(z,y)\varrho_2(y)\dd
y\right]^2\dd z\leq\nonumber\\
&&\int_{\mbb{R}^K}{\Phi^2(z,R^{(l)})\over\phi^2(R^{(l)})}\dd
zL^2_{\Psi}\left(\int_{\mbb{R}^K}\left|\varrho_1(y)-\varrho_2(y)\right|\dd
y\right)^2\leq
L_{\Phi}^2L_{\Psi}^2\int_{\mbb{R}^K}\left|\varrho_1(z)-\varrho_2(z)\right|^2\dd
z\nonumber
\end{eqnarray}

Finally, since $\varrho_2(R)$ is {\it globally} Lipschitz continuous
(there is a common Lipschitz constant $L_R$ for all $R(\omega)$),
\begin{eqnarray}
\lefteqn{\int_{B(\vec{k})}|\varrho_2(R)(z)-g_k|^2\dd z\leq
2\int_{B(\vec{k})}|\varrho_2(R)(z)-\varrho_2(R)(z_n)|^2\dd
z+}\nonumber\\&&2\mathrm{vol}(B(\vec{k}))|\varrho_2(R)(z_n)-g_n|^2\leq
2L^2_R\mathrm{vol}(B(\vec{k}))\mathrm{diam}(B(\vec{k}))^2+\nonumber\\&&2\mathrm{vol}(B(\vec{k}))|\varrho_2(R)(z_n)-g_n|^2\nonumber
\end{eqnarray}
Then,
\begin{eqnarray}
\lefteqn{\mathit{err}_n\leq\epsilon+
2L_x(1+r(t))|x_1-x_2|+2L_{\varrho}L_{\Phi}^2L_{\Psi}^2\int_{\mbb{R}^K}\left|\varrho_1(z)-\varrho_2(z)\right|^2\dd
z+}\nonumber\\&&2L_{\varrho}L^2_R\mathrm{vol}(B(\vec{k}))\mathrm{diam}(B(\vec{k}))^2+2L_xn^{-2}J_1\left(\int_{\mbb{R}}\left(f_X(x)\right)^{1\over
3}\dd x\right)^3+\nonumber\\
&&
4L_{\varrho}^2\mbf{E}\left[\min_{\{Q\in\mcal{S}_Q|\#Q=n\}}\sum_{\vec{k}\in\hat{K}_n}\mathrm{vol}(B(\vec{k}))|\varrho(R)(z_k)-g_k|^2\right]+2L_{\varrho}\int_{\mathfrak{B}^c_n}\varrho(R)^2(z)\dd
z.\nonumber
\end{eqnarray}
The sign ''$\sum$'' can be swapped with ''$\min$'' since for each 2
elements of the sum, the minimum is calculated with respect to 2
different arguments. Zador Theorem and assumption on $B(\vec{k})$
imply
\begin{eqnarray}
\lefteqn{\mathit{err}_n\leq\epsilon+
2L_x(1+r(t))^2|x_1-x_2|^2+2L_{\varrho}L_{\Phi}^2L_{\Psi}^2\int_{\mbb{R}^K}\left|\varrho_1(z)-\varrho_2(z)\right|^2\dd
z}\nonumber\\
&&L_xn^{-2}J_1\cdot\left(\int_{\mbb{R}}f_X(x)^{1\over 3}\dd x\right)^3+2L_{\varrho}L_R^2\left({M_0\over n}\right)^K\cdot\sqrt{K}\left({M_0\over n}\right)+\nonumber\\&&\quad 4L_{\varrho}\left({M_0\over n}\right)^Kn^{-{2\over N}}J_N\cdot\sum_{\vec{k}\in\hat{K}_n}\left(\int_{\mbb{R}^N}f_{z_k}(x)^{N\over N+2}\dd x\right)^{N+2\over N}+2L_{\varrho}\int_{\mathfrak{B}^c_n}\varrho(R)^2(z)\dd z\nonumber\\
&&\leq
\epsilon+2L_x(1+r(t))^2|x_1-x_2|^2+2L_{\varrho}L_{\Phi}^2L_{\Psi}^2\int_{\mbb{R}^K}\left|\varrho_1(z)-\varrho_2(z)\right|^2\dd
z\nonumber\\
&&L_xn^{-2}J_1\cdot\int_{\mbb{R}^K}\left(f_X(x)^{1\over 3}\dd
x\right)^3+2L_{\varrho}L_R^2\sqrt{K}\left({M_0\over
n}\right)^{K+1}+\nonumber\\&& \quad 4L_{\varrho}M_0^Kn^{-{2\over
N}}J_N\left(\int_{\mbb{R}^N}f^{max}_{z_k}(x)^{N\over N+2}\dd
x\right)^{N+2\over
N}+2L_{\varrho}a_z\int_{[M_0,+\infty)^N}|x|^{-p}\dd x\nonumber
\end{eqnarray}
The calculation of the constant $v_N$ is left to the Appendix.
\koniec
\begin{uwaga}\label{rem:minimumExists}
The minimum with respect to $p_i$s in formula \ref{eq:zadorInAction}
$\int_{\mbb{R}}(((\phi^{(e)}_{p_1}\ast...\ast\phi^{(e)}_{p_N}))(x))^{1\over
2}\dd x$ always exists. Let us show it in the case of $N=2$. The
other cases are analogous.

We have
\begin{eqnarray}F(p_1,p_2)&\bydefi&\int_{\mbb{R}}\left(((\phi^{(e)}_{p_1}\ast...\ast\phi^{(e)}_{p_2}))(x)\right)^{1\over
2}\dd x\nonumber\\&=&\int_{\mbb{R}}\left(\int_{\mbb{R}}{1\over
p_1p_2}\phi^{(e)}\left({x-z\over p_1}\right)\phi^{(e)}\left({z\over
p_2}\right)\dd z\right)^{1\over 2}\dd x
\end{eqnarray}
and obviously, by the
definition of $\phi^{(e)}_p$, $F(0,p_2)=F(p_1,0)=0$. It is
straightforward that $F$ is continuous on the compact set
$[\bar{p}_1,x_1]\times[\bar{p}_2,x_2]$ and
$f|_{[\bar{p}_1,x_1]\times[\bar{p}_2,x_2]}$ attains the maximum.

To show that $F$ has maximum on $[0,x_1]\times[0,x_2]$ it is
sufficient to prove that $F$ has negative (and continuous) partial
derivatives on $(0,\bar{p}_1)\times(0,\bar{p}_2)$, for some
$\bar{p}_1$ and $\bar{p}_2$. Let us concentrate on the partial
derivative with respect to $p_1$. Firstly, we majorize $\phi^{(e)}$
by an integrable function $\hat{\phi}^{(e)}$ with the monotone tails
in the sense that for every $i$
$\hat{\phi}^{(e)}(x_1,\dots,x_{i-1},\cdot,x_{i+1},\dots,x_N)$ is
increasing for sufficiently large negative arguments and decreasing
for sufficiently large positive ones. It is always possible for the
bounded density function. For some mild conditions we can swap the derivative
and integrals.
\begin{eqnarray}
\lefteqn{{\partial F(p_1,p_2)\over\partial p_1}=\int_{\mbb{R}}{1\over 2}\left(\int_{\mbb{R}}{1\over p_1p_2}\hat{\phi}^{(e)}\left({x-z\over p_1}\right)\hat{\phi}^{(e)}\left({z\over p_2}\right)\dd z\right)^{-{1\over 2}}\cdot}\nonumber\\
&\cdot&\int_{\mbb{R}}{1\over p_2}\left[-{1\over
p_1^2}\hat{\phi}^{(e)}\left(x-z\over
p_1\right)-\nabla\hat{\phi}^{(e)}\left({x-z\over
p_1}\right){x-z\over p_1^3}\right]\hat{\phi}^{(e)}\left(z\over
p_2\right)\dd z\dd x.\nonumber
\end{eqnarray}
The expression in the square brackets is negative for sufficiently small
$p_1$. The first component is always negative. If $x_i-z_i\geq 0$
than for $p_1$ large enough $\left({x_i-z_i\over
p_1}\right)\nabla\hat{\phi}^{(e)}_i\left({x-z\over p_1}\right)\leq
0$. On the other hand, $x_i-z_i<0$ implies $$\left({x_i-z_i\over
p_1}\right)\nabla\hat{\phi}^{(e)}_i\left({x-z\over p_1}\right)>0.$$
Thus, on $[0,\bar{p}_1]\times[0,\bar{p}_2]$, $F$ is bounded by a
function that has the maximum value. This completes the proof.
\end{uwaga}
A direct consequence of lemma \ref{lemat:lipschVT} is the following
estimate of the approximation error of the algorithm.
\begin{tw}\label{tw:ApprError}
For $u_T(\cdot,z)$ $L_u$-Lipschitz continuous and $u_T(\cdot,\cdot)$
bounded by $L_u^{(b)}$,
\begin{eqnarray}
\lefteqn{|V(0,x,\varrho)-\widehat{V}(0,\widehat{x},\widehat{\varrho})|^2<\delta\frac{1-\delta^T}{1-\delta}C_{n,M_0}+}\nonumber\\
&&\qquad\qquad\qquad\qquad\qquad L^{(1)}_T|x-\widehat{x}|^2+L^{(2)}_T\int_{\mbb{R}^K}|\varrho(z)-\widehat{\varrho}(z)|^2\dd
z,
\end{eqnarray}
with
\begin{eqnarray}
C_{n,M_0}&\bydefi&\left(L_un^{-2}J_1\cdot\left(\int_{\mbb{R}}f_X(x)^{1\over
3}\dd x\right)^3+2L_u^{(b)}L_R^2\sqrt{K}\left({M_0\over
n}\right)^{K+1}+\right.\nonumber\\&&
\left.4L_u^{(b)}M_0^Kn^{-{2\over
N}}J_N\left(\int_{\mbb{R}^N}f^{max}_{z_k}(x)^{N\over N+2}\dd
x\right)^{N+2\over N}+2L_u^{(b)}a_zv_N{M_0^{N-p}\over
p-1}\right),\nonumber\\
L_T^{(1)}&\bydefi&2^{T+1}\delta^TL^2_u\prod_{t=1}^T(1+r(t)),\nonumber\\
L^{(2)}_T&\bydefi&2^{T+1}\delta^T\left(L_u^{(b)}\right)^2L_{\Phi}^{2T}L_{\Psi}^{2T}.\nonumber
\end{eqnarray}
\end{tw}
\dowod Let us suppose that the projections ''$\widehat{\cdot}$''
realizes the minimal error over the set of projection sets and
$V(t+1,\cdot,\cdot)$ is $(\epsilon,L_x,L_{\varrho})$-Lipschitz
continuous. In this case, applying lemma \ref{lemat:lipschVT},
\begin{eqnarray}
\lefteqn{|V(t,x,\varrho)-\widehat{V}(t,\widehat{x},\widehat{\varrho})|^2}\nonumber\\
&\leq&\Big(\sup_{(c,\pi)\in\mcal{A}_t(x)\cup\mcal{A}_t(\widehat{x})}\Big\{u(c)+\delta\bar{\mbf{E}}V(t+1,x(R),\varrho(R))-\nonumber\\
&&\qquad\qquad\qquad\qquad\qquad\left.\left.+u(c)-\delta\bar{\mbf{E}}\widehat{V}(t+1,\widehat{x(R)},\widehat{\varrho(R)})\right\}\right)^2\nonumber\\
&\leq&\sup_{(c,\pi)\in\mcal{A}_t(x)\cup\mcal{A}_t(\widehat{x})}\left\{\delta\bar{\mbf{E}}\left|V(t+1,x(R),\varrho(R))-\widehat{V}(t+1,\widehat{x(R)},\widehat{\varrho(R)})\right|^2\right\}\nonumber\\
&=&\delta\epsilon+\delta C_{n,M_0}+2\delta
L_x(1+r(t))^2|x-\widehat{x}|^2+2\delta
L_{\varrho}L_{\Phi}^2L_{\Psi}^2\int_{\mbb{R}^K}|\varrho(z)-\widehat{\varrho}(z)|^2\dd
z.\nonumber
\end{eqnarray}
Thus, $V(t,\cdot,\cdot)$ is $(\delta\epsilon+\delta C_{n,M_0},2\delta
L_x(1+r(t))^2,2\delta
L_{\varrho}L_{\Phi}^2L_{\Psi}^2)$-Lipschitz continuous. By induction the proof is completed.\koniec
\section{Numerical illustration}
We show the performance of the algorithm for an arbitrary
parametrization of the model. We consider the market for only one
risky asset with the stochastic volatility driven by the one dimensional
process $y$. Even in this case, the computation of the optimal
consumption / investment paths is very much time consuming for the
algorithm implemented on a regular computer\footnote{1.6MHz dual
core processor with 2GB RAM.}.

We assume that the dynamics of the market specified in general
terms in the subsection \ref{sec:intro} is given by the following
system of the three equations:
\begin{eqnarray}
S_0(t+1)&=&(1+0.03)^{t+1},\nonumber\\
S_1(t+1)&=&S_1(t)\exp\left\{0.05+0.25(1-0.75\exp\{-Y(t+1)^2\})\epsilon(t+1)\right\},\nonumber\\ S_1(0)&=&6.00,\nonumber\\
Y(t+1)&=&0.80\cdot Y(t)+0.20\xi(t+1),\nonumber\\
Y(0)&=&1.50.\label{eq:exampleY}
\end{eqnarray}
The risk-free interest rate (the bank account interest) is equal to 3\%.
We assume that the mean log rate of return from the risky asset is
5\%. The volatility of the risky asset can vary a lot in the model.
It can be as large as almost 16\% for $Y(t)=0.8$ and as low as 6\% if
$Y(t)=0.0$. For the parametrization in the equation
\ref{eq:exampleY} trajectories of $Y$ usually oscillates within -0.8 and 0.8 bounds.

The variability of $Y$ is visibly reflected into the volatility
of $S_1$. A sample trajectory of $S_1$ is shown on figure
\ref{exPrices}.
\wstawrysunekRb{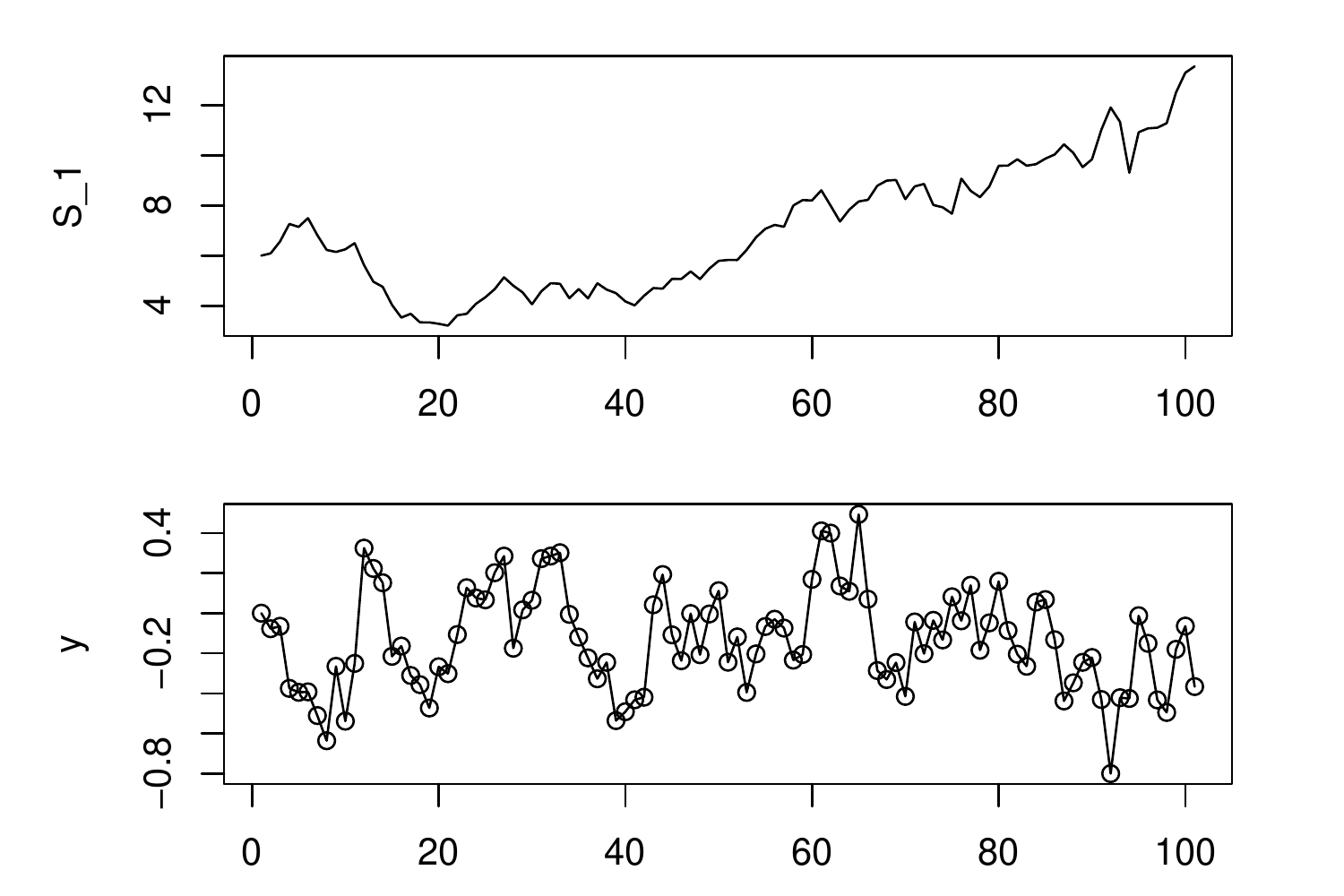}{exPrices}{A trajectory
of the price process and underlying parameter $y$, i.e. trajectory
of the stochastic volatility.}{own calculations}{}{}{}

The utility functions meeting the requirements of boundedness and
Lipschitz continuity demanded in the assumptions of the theorem
\ref{tw:ApprError} (in fact the lemma \ref{lemat:lipschVT}) were
chosen to be the following:
\begin{eqnarray}
u_T(y,x)&=&\left(2\exp(-y)x\over 1+2\exp(-y)x\right)^{0.5}\
\mathrm{and}\ u(x)=0.25\left(x\over 2+x\right)^{0.5}.
\end{eqnarray}
This are the utility function of the so-called IRRA type, i.e. the
\textit{increasing relative risk aversion}.

The numerical algorithm to calculate the approximate optimal
consumption / investment strategies was based on the backwards
dynamic programming equation \ref{eq:approxDynProg}. We compute
$\widehat{V}$ recursively for 10 points of time with the time
horizon $T=10$. At each point of time equation
\ref{eq:approxDynProg} give us the estimates of the consumption $c$ and the
investment $\pi$ maximizing his joint utility.

Keeping in mind the complexity of the calculation of
$\widehat{V}(t,\widehat{x},\widehat{\varrho})$ based on
$\widehat{V}(t+1,\cdot,\cdot)$ we have to chose a limited number of nodes for the portfolio value $x$ and the beliefs $\varrho$.
To make the calculation tractable, we assume that the quantization
set of the portfolio process $X$ is
$Q^X\bydefi\{0.00,0.25,0.50,\dots,9.75,10.00\}$. The investor's beliefs
described by the unnormalized densities are projected on the appropriate
quantization.

Since the minimal error in the approximate algorithm described by the recursion \ref{eq:approxDynProg} is realized for the optimally chosen quantization set we apply the stochastic gradient descent algorithm to find the optimal structure (compare \citet{Pages2}). We start with the initial arbitrary set of the densities $\mathcal{Q}_n^0$ evaluated at the points from $Z_n\colon=\{-1.50,\,-1.45,\,-1.40,\dots,\,1.45,\,1.50\}$. We choose the parameters -- the mean and the standard deviation -- to be the equidistant point on the real line, i.e. the mean took values in the set $\{-1.50,\,-1.25,\,\dots,\,1.50\}$ and the standard deviation in the set $\{0.1,\,0.3,\,0.5,\,0.7,\,0.9\}$. The densities are presented on figure \ref{q0qopt}, top panel.
\wstawrysunekR{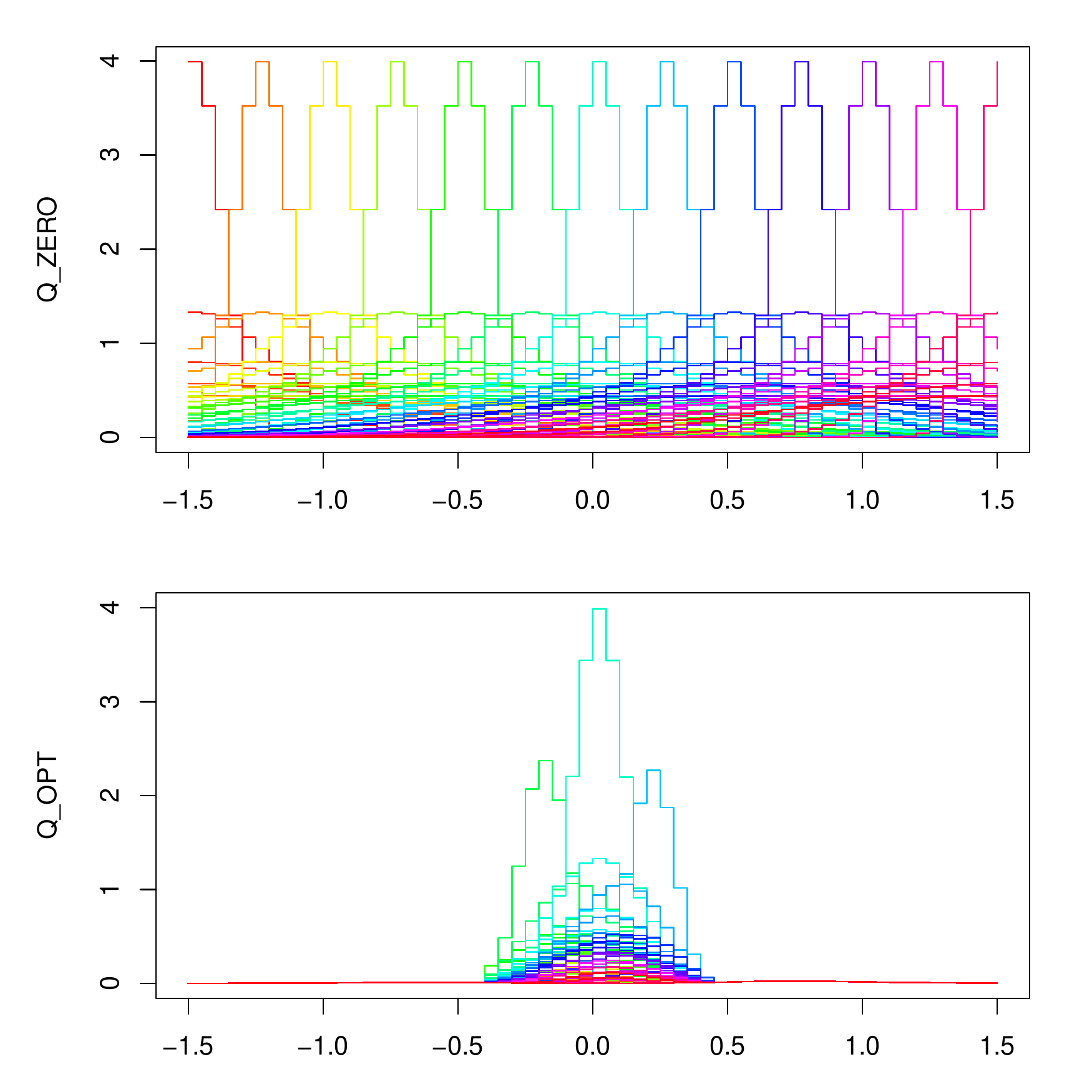}{q0qopt}{The initial quantization set $\mathcal{Q}^0_n$ (top graph) and the approximation of the optimal quantization set in the stochastic gradient descent methods in 500 steps (bottom graph)}{own calculations}{}{}{}

\wstawrysunekRhalf{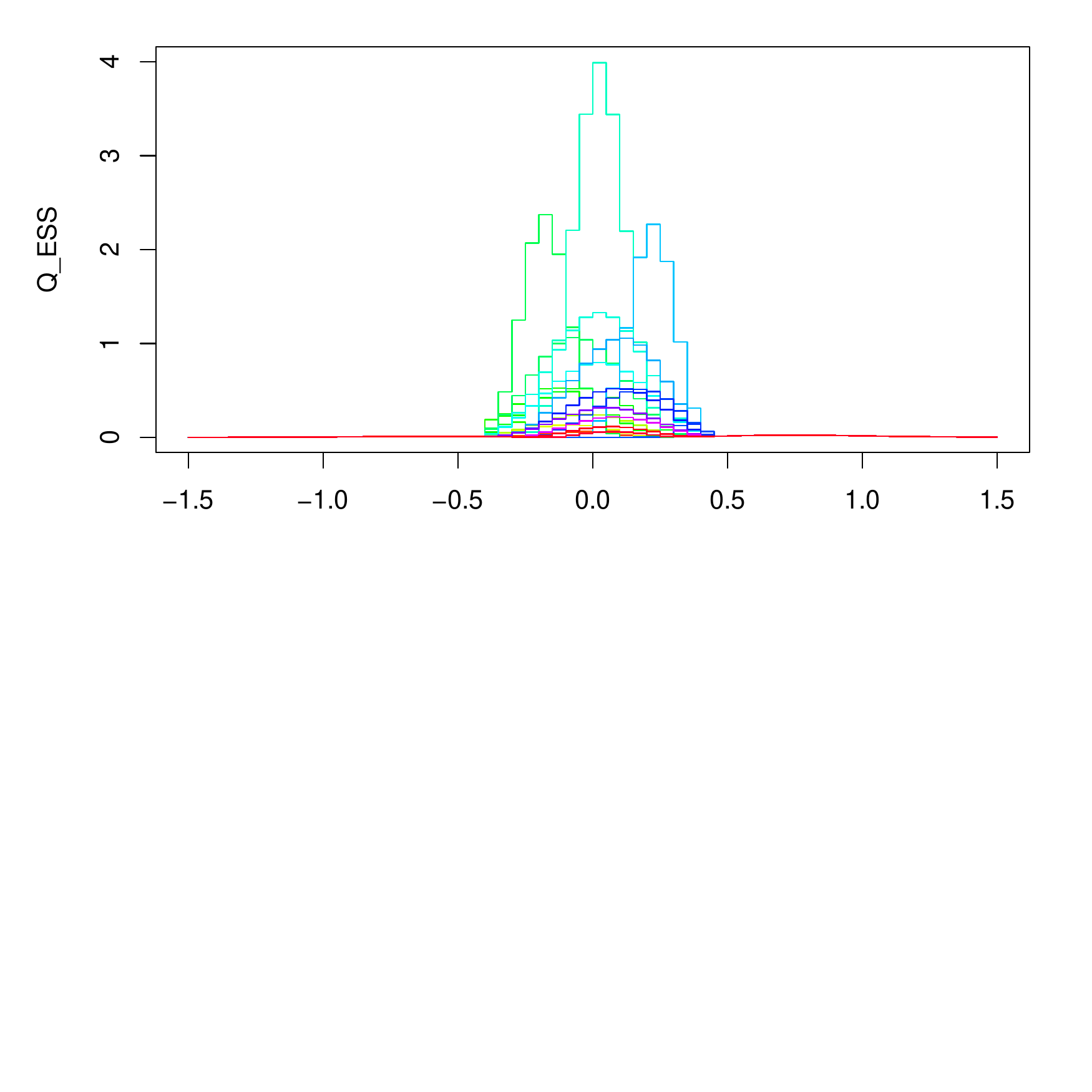}{qess}{The selection of the essential elements of the quantization set $\mcal{Q}^{500}_n$ (25 out of 65 elements)}{own calculations}{}{}{}

The procedure requires to construct the incremental function $H$ which defines the corrections of $\mcal{Q}^0_n$ leading in the limit to the optimal quantization set. The incremental function is derived by the formal differentiation under $\bar{\mbf{E}}$ of the distortion function $D_n(Q)$ for a given realization $\hat{r}$ of return $R^l$. Namely, in the special case of 1-dimensional process $Y$ any quantization set $\mcal{Q}$ can be represented by the matrix from $\mcal{M}^{n\times m_n}$ where $(k,m)$ element is defined as the value of the $k$th element of the set $\mcal{Q}$ evaluated at the $m$th argument from $Z_n$. Then, the function $H\colon\mcal{M}^{n\times m_n}\times\mbb{R}\to\mcal{M}^{n\times m_n}$ is defined as
$$H_{k,m}(\mcal{Q},\hat{r})\colon={\partial\over\partial\hat{r}}\varrho(z_k,\hat{r})\left(\varrho(z_k,\hat{r})-g_m(k)\right).$$
Since in the example the densities $\phi$ and $\psi$ in the example are smooth then the derivative of $\varrho$ is well-defined. Then for a given sequence $\{\beta_i\}_{i\in\mbb{N}}$ satisfying the usual regularity conditions ($\sum\beta_i=\infty$ and $\sum\beta^2_i<\infty$) the following sequence of the quantization sets converge to the optimal one (see \citet{Shapiro} for convergence results): $$\mcal{Q}_n^{m+1}=\max\left\{\mcal{Q}_n^m+\beta_{m+1}H(\mcal{Q}_n^m,R^l(\omega_{m+1})),0\right\}.$$

Applying this recursive procedure for $\mcal{Q}_n^m$ with 500 iterations of $m$ we obtain the modified set $\mcal{Q}^{500}_n$ which is graphically presented on figure \ref{q0qopt}, bottom panel. Firstly, one crucial advantage of $\mcal{Q}_n^{500}$ is the reduction of the domain of the densities from the initial $[-1.5,1.5]$ to $[-0.5,0.5]$ where they are essentially different from 0. This decreases the dimensionality of the numerical problem since the same number of points $Z_n$ can be denser allocated on the smaller interval $[-0.5,0.5]$. Secondly, we can drop the densities from $\mcal{Q}_n^{500}$ which are very similar to each other and we take into account only a subset $\mcal{Q}_n^{ess}\subset\mcal{Q}_n^{500}$ if the densities. Namely, we apply the following procedure:
\begin{enumerate}
\item $\mcal{Q}_n^{ess}\colon=\mcal{Q}_n^{500}$\verb+;+ \verb+eps=0.1;+
\item \verb@REPEAT i@
\item \verb@Chose a pair of densities@ $q_1\in\bar{Q}^{ess}$ \verb+and+ $q_2\in\bar{Q}^{ess}$\verb+;+
\item \verb+IF+ $\frac{\sum_{z\in Z_n}|q_1(z)-q_2(z)|}{\sum_{z\in Z_n}|q_1(z)|}<$\verb+eps THEN+ $\mcal{Q}_n^{ess}\colon=\mcal{Q}_n^{ess}/\{q_2\}$\verb+;+
\item \verb@i:=i+1@\verb+;+
\item \verb@UNTIL i==5000;@
\end{enumerate}
the application of which gives the quantization set consisting of 25 unnormalized densities.

It should be underlined that functions $\varrho$ representing the beliefs about $Y$, described by recursion
\ref{eq:recursive_filter}, are not necessarily densities. They are
unnormalized densities. An example of the evolution of beliefs is presented on the figure
\ref{evolBeliefs}.

\wstawrysunekRb{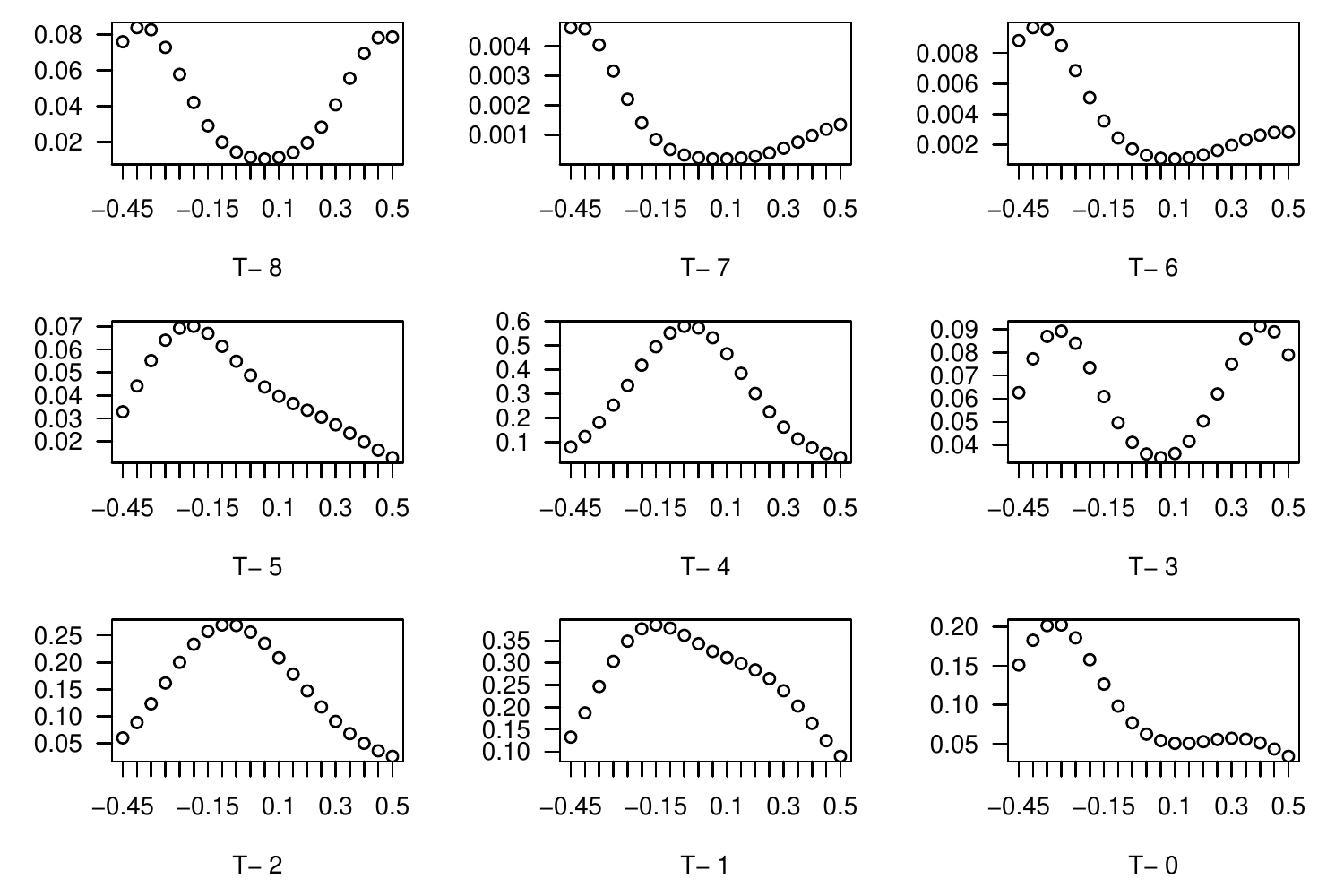}{evolBeliefs}{Evolution of beliefs
with prior at time $T-9$ equal to the normal distribution with mean
$0.1$ and standard deviation $0.2$}{own calculations}{Each box shows
a density which is recursively obtained using equation
\ref{eq:recursive_filter}, i.e. the density on the box $T-n$ is
obtained from the formula \ref{eq:recursive_filter} with
$\varrho_{T-n-1}$ equal to the density from the box $T-n-1$ and
$\varrho_{T-9}\equiv\phi_{0.1,0.2}$.}{Note}{}

The number of nodes of the portfolio values, which is 41, multiplied
by the number of the densities in the density quantization set equal
to 25 gives 1025 arguments of $\widehat{V}(t,\cdot,\cdot)$.

Searching for the optimal consumption and investment for a given
point of time $t$, the current wealth $\bar{x}$ and the beliefs about the
unobserved factors driving the volatility on the market, we
discretize the space of investors decisions. We were looking for
the maximum expected value of the value function
$\widehat{V}(t+1,x^{c,\pi}(t+1),\varrho_{t+1})$ among
$c\in\{0.00\bar{x},0.05\bar{x},0.10\bar{x},\dots,0.95\bar{x},1.00\bar{x}\}$
and
$\pi\in\{0.00\bar{x},0.05\bar{x},0.10\bar{x},\dots,0.95\bar{x},1.00\bar{x}\}$.

\wstawrysunekRb{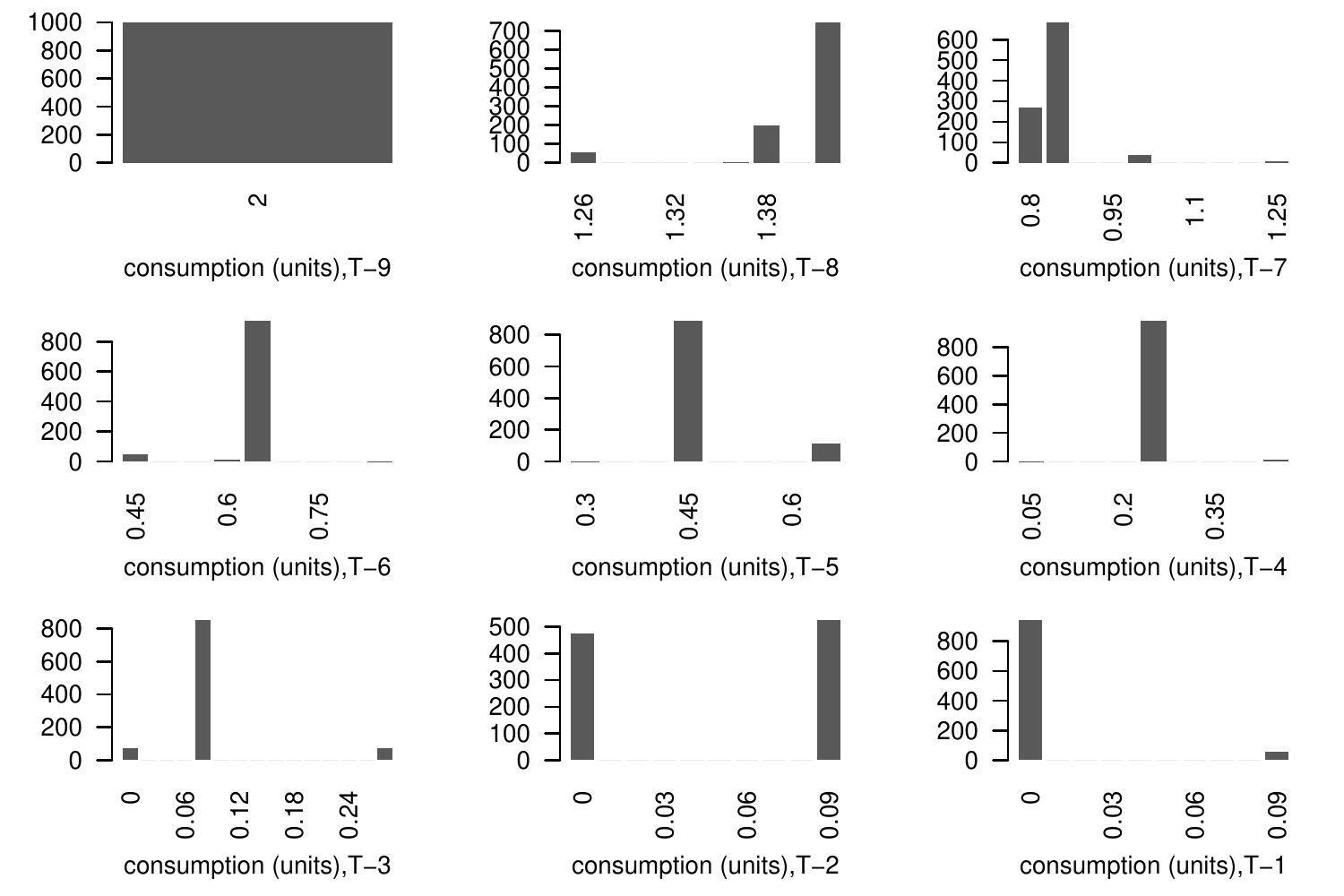}{optCons}{Distribution of the
optimal consumption for different points of time}{own
calculation}{histograms for 1000 simulation of the risky asset
dynamics. $T-n$ means \textit{$n$ time periods before the time
horizon $T$}.}{Note}{} The four main components of the solution of
the numerical optimization were presented: the consumption, the investment
in the risky asset and the risk-free asset and the value of the optimal portfolio. Assuming
that the investor possesses 6 units of the initial capital we
generated his wealth paths along the estimated consumption and
investment policies. We perform 1000 simulations of the risky asset
prices and we construct the corresponding consumption and
investment paths according to the approximate optimal policy.

The results of the simulation suggest that the optimally behaving
investor consume about half of the initial wealth during the first
two periods and then gradually decreases the consumption rate (see figure \ref{optCons}). In the period $t=6$ the investor is left with less then 1 units of wealth. In the last three periods he consumes only about 0.1 units of his
portfolio, almost irrespective of the market situation (i.e.
independent of the evolution of the risky asset prices).

\wstawrysunekRb{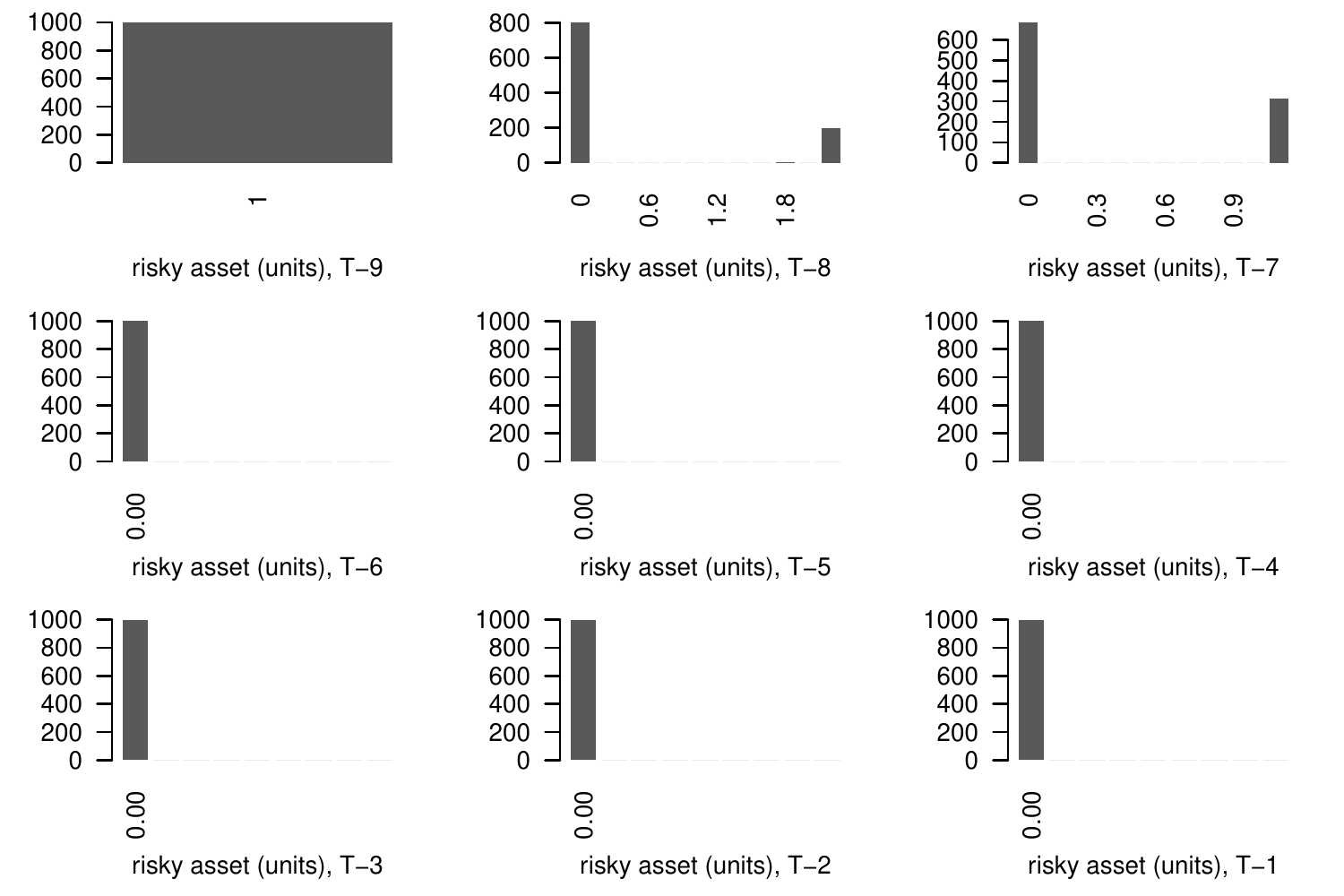}{optRiskyAsset}{Distribution of
the optimal investment into the risky asset for different points of
time}{own calculation}{histograms for 1000 simulation of the risky
asset dynamics. $T-n$ means \textit{$n$ time periods before the time
horizon $T$}.}{Note}{} Conversely, his investment decisions are much
more dependent on the market trends and on the time of decision
making. His stronger beliefs about unfavorable volatility (about
$y$) decrease his propensity to invest into the risky asset. The investor allocates quite a small part of his wealth in the risky project $S_1$ only in the first two periods (see figure \ref{optRiskyAsset}). The
closer to the terminal time $T$ the more he is inclined to allocate
his current wealth into the risk-free asset $S_0$ (see figure \ref{optRiskFreeAsset}). As the figure
\ref{optPortf} indicates, the terminal wealth reaches almost 0 units.
\wstawrysunekRb{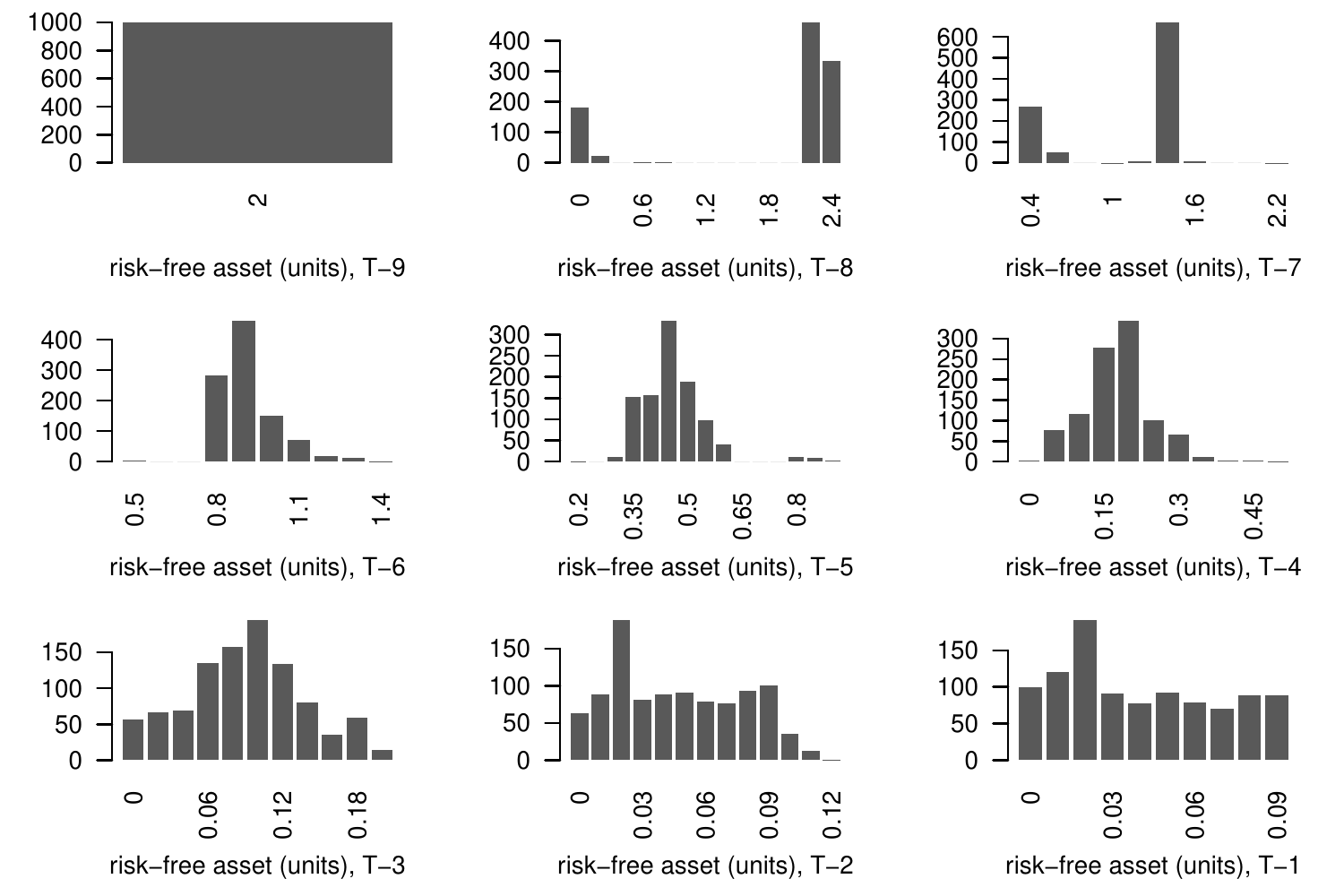}{optRiskFreeAsset}{Distribution of
the optimal investment into the risk-free asset for different points of
time}{own calculation}{histograms for 1000 simulation of the risky
asset dynamics. $T-n$ means \textit{$n$ time periods before the time
horizon $T$}.}{Note}{}

\wstawrysunekRb{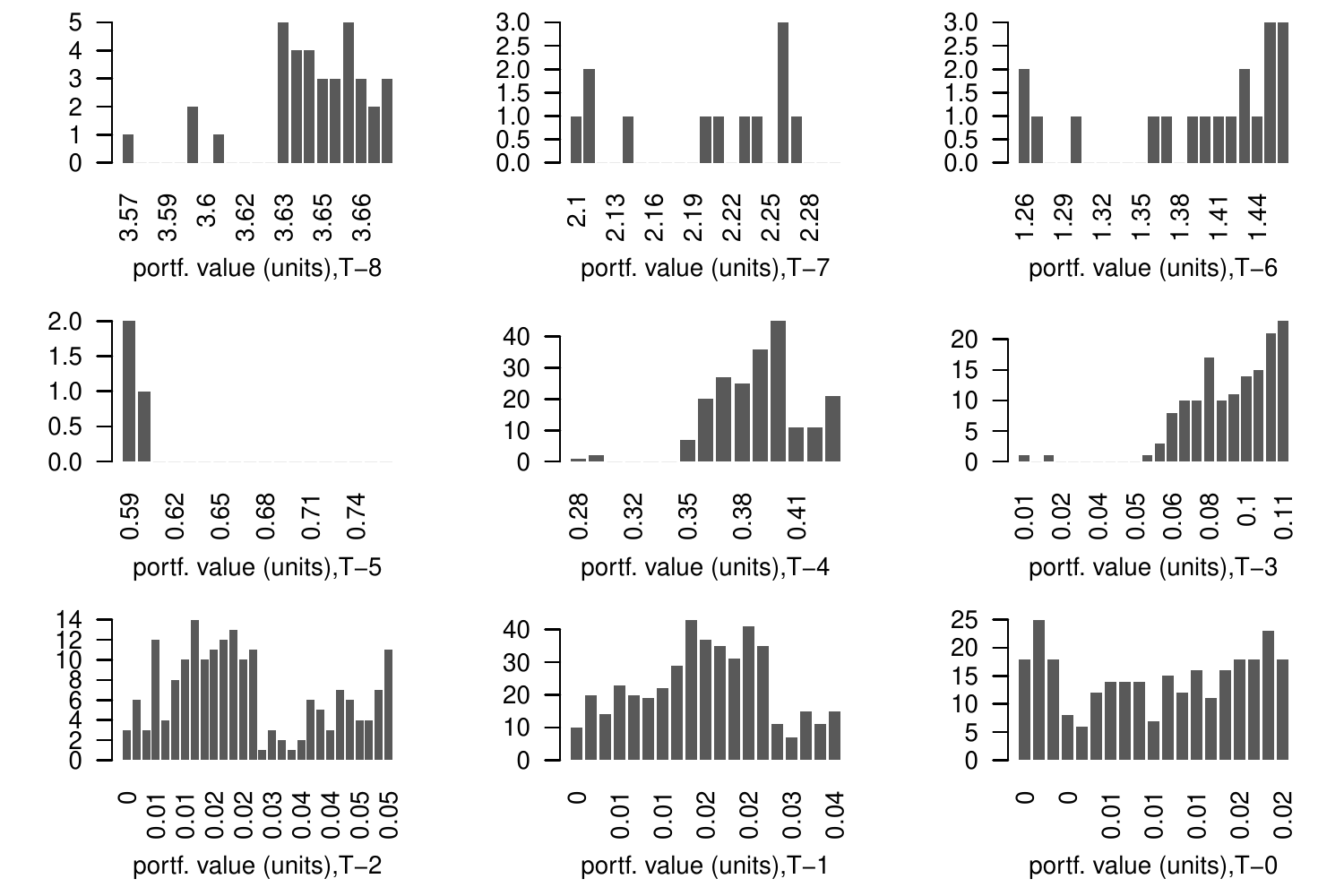}{optPortf}{Distribution of the optimal
value of portfolio for different points of time}{own
calculation}{histograms for 1000 simulation of the risky asset
dynamics. $T-n$ means \textit{$n$ time periods before the time
horizon $T$}.}{Note}{}

Even though the set of beliefs seems to be only roughly approximated
by 30 density functions, the simulation reflects the behavior of a
rational investor in a quite reasonable and intuitive way. However, the
computational burden of the numerical procedure makes it difficult
to support investment decisions on a market for multiple assets
unless the complexity of the stochastic volatility part of the model
is substantially reduced. In fact the minimal error of the estimate of the value function $V$ that can be attained in the numerical example and given by the constant $C_{n.M_0}$ multiplied by the factor $\delta\frac{1-\delta^T}{1-\delta}$ (see theorem \ref{tw:ApprError}) is equal to 0.81 which is about 12\% of the utility of the initial wealth.

However, the proposed quantization framework follows the promising avenue of research on the portfolio optimization under the partial
observation of the stochastic volatility. It shows that the nonlinear
filtering can be tractable even in the dynamic programming setting.
From the financial point of view, the incomplete information is a very
important driver of the market dynamics and its influence on the
economic agents' optimal choice is worse studying. Certainly, a more
efficient numerical implementation of the algorithm is required.
\section{Conclusions}
We solved the dynamic optimal consumption/investment strategy under
partial observation of economic factors determining asset price
movements by means of the numerical method. The special variation of
the quantization technique reduced the originally infinite
dimensional problem to the finite state space exercise. We successfully developed the algorithm to solve the consumption / investment problem on the stylized market for 2 tradable assets and 1 unhedgable economic factor.

The extension of the paper can be devoted to a study of the
stability of the algorithm solving numerically the optimal portfolio
choice with consumption. The first important question remains how
the results of the simulation change between two different sets of
randomly generated asset returns. The second one could be related to
the size of quantization set that would in practice give an assumed
accuracy of numerical optimal strategies.

We transformed the partial observation to the full observation
problem, assuming that the shocks affecting prices and economic
factors are independents. However, existing methods of stochastic
volatility estimation \citep{Durham} do not need shocks to be
orthogonal. It would be interesting to develop a model of
consumption/investment with imperfect observation and correlation of
volatility and idiosyncratic shocks in the asset price movements.
\appendix
\section{Filters}\label{app:filter}
The following lemma is the most crucial in obtaining filters of
partially observed processes. It states that $Y$s and $R_i^l$s are
independent in $(\mcal{F}^{lR},\bar{\mbf{P}})$. After
disentangling $Y$ from $R^l$, the evolution of the variable $Y$ can be reformulated to a
recursive equation of measures on $(\Omega,\mbb{F}^{lR})$.
\begin{lemat}
The processes $Y$ and $R^l$ are independent in $\mcal{F}^{lR}$ under
$\bar{\mbf{P}}$.
\end{lemat}
\dowod Let us take two Borel functions $f_1$ and $f_2$. Then
\begin{eqnarray}
\barcondE{f_1(Y(t))f_2(R^l(t))}{F}{lR}{t-1}&=&{\condE{\Lambda_t^{-1}f_1(Y(t))f_2(R^l(t))}{F}{lR}{t-1}\over\condE{\Lambda_t^{-1}}{F}{lR}{t-1}}\nonumber\\
&=&{\condE{\lambda_t^{-1}f_1(Y(t))f_2(R^l(t))}{F}{lR}{t-1}\over\condE{\lambda_t^{-1}}{F}{lR}{t-1}}\label{eq:independence_bayes}
\end{eqnarray}
since $\Lambda^{-1}_t=\lambda^{-1}_t\Lambda^{-1}_{t-1}$ and
$\Lambda^{-1}_{t-1}$ is $\mcal{F}^Y_{t-1}$-measurable. For
brevity $\tilde{y}\bydefi Y(t-1)$. Let us take the denominator in
\ref{eq:independence_bayes} and
\begin{eqnarray}
\lefteqn{\condE{\lambda_t^{-1}}{F}{lR}{t-1}=}\nonumber\\
&&=\condE{{\det(\sigma(Y(t)))\phi(R^{(l)}(t))\over\phi\Big(\sigma^{-1}(Y(t))[R^{(l)}(t)-\mu(Y(t))]\Big)}{\det(\sigma_Y)\psi(Y(t))\over\psi\Big(\sigma_Y^{-1}[Y(t)-\tilde{y}-\alpha(\bar{y}-\tilde{y})]\Big)}}{F}{lR}{t-1}\nonumber\\
&&=\condE{{\phi\big(\mu(Y(t))+\sigma(Y(t))\epsilon(t)\big)\over\det(\sigma^{-1}(Y(t)))\phi\left(\epsilon(t)\right)}{\psi\big(\tilde{y}-\alpha(\bar{y}-\tilde{y})+\sigma_Y\xi(t)\big)\over\det(\sigma^{-1}_Y)\psi(\xi(t))}}{F}{lR}{t-1}.\nonumber
\end{eqnarray}
Put
\begin{eqnarray}
p_1\bydefi{\psi\big(\tilde{y}-\alpha(\bar{y}-\tilde{y})+\sigma_Y\xi(t)\big)\over\det(\sigma^{-1}_Y)\psi(\xi(t))}.\nonumber
\end{eqnarray}
Then
\begin{eqnarray}
\condE{\lambda_t^{-1}}{F}{lR}{t-1}=\condE{p_1\mbf{E}\left[{{\phi\big(\mu(Y(t))+\sigma(Y(t))\epsilon(t)\big)\over\det(\sigma^{-1}(Y(t)))\phi\left(\epsilon(t)\right)}}\Big|\sigma(\mcal{F}^{lR}_{t-1},\mcal{F}^Y_{t})\right]}{F}{lR}{t-1}.\nonumber
\end{eqnarray}
The variable $\epsilon(t)$ is independent of $\mcal{F}_{t-1}^{lR}$
so we can integrate with respect to the density $\phi$
\begin{eqnarray}
\condE{\lambda_t^{-1}}{F}{lR}{t-1}=\condE{p_1\int_{\mbb{R}^N}{\phi\big(\mu(Y(t))+\sigma(Y(t))\hat{\epsilon}\big)\over\det(\sigma^{-1}(Y(t)))\phi\left(\hat{\epsilon}\right)}\phi(\hat{\epsilon})\dd\hat{\epsilon}}{F}{lR}{t-1}.\nonumber
\end{eqnarray}
By changing variables with the diffeomorphism
$\Theta(\hat{\epsilon})=\sigma^{-1}(Y(t))\hat{\epsilon}$, with a
given fixed $\omega$, as follows
$\int_{\Theta(\mbb{R}^N)}G(\hat{\epsilon})\dd\hat{\epsilon}=\int_{\mbb{R}^N}(\Theta\circ
G)(\hat{\epsilon})\det(\Theta'(\hat{\epsilon}))\dd \hat{\epsilon}$
we get
\begin{eqnarray}
\lefteqn{\condE{\lambda_t^{-1}}{F}{lR}{t-1}=\condE{p_1\int_{\mbb{R}^N}\phi\big(\mu(Y(t))+\hat{\epsilon}\big)\dd\hat{\epsilon}}{F}{lR}{t-1}}\nonumber\\
&&=\condE{\mbf{E}\left[{\psi\big(\tilde{y}-\alpha(\bar{y}-\tilde{y})+\sigma_Y\xi(t)\big)\over\det(\sigma^{-1}_Y)\psi(\xi(t))}\Big|\sigma(\mcal{F}^{lR}_{t-1},\mcal{F}^Y_{t-1})\right]}{F}{lR}{t-1}\nonumber\\
&&=\condE{\int_{\mbb{R}^K}{\psi\big(\tilde{y}-\alpha(\bar{y}-\tilde{y})+\sigma_Y\hat{\xi}\big)\over\det(\sigma^{-1}_Y)\psi(\hat{\xi})}\psi(\hat{\xi})\dd\hat{\xi}}{F}{lR}{t-1}\equiv
1,\nonumber
\end{eqnarray}
observing that $\xi(t)$ is independent of $\mcal{F}^{lR}_{t-1}$.

Let us take
\begin{eqnarray}
\lefteqn{\condE{\lambda_t^{-1}f_1(Y(t))f_2(R^l(t))}{F}{lR}{t-1}=}\nonumber\\
&&=\condE{p_1f_1(Y(t))f_2(R^l(t)){\phi\big(\mu(Y(t)))+\sigma(Y(t))\epsilon(t)\big)\over\det(\sigma^{-1}(Y(t)))\phi\left(\epsilon(t)\right)}}{F}{lR}{t-1}\nonumber\\
&&=\condE{p_1f_1(Y(t))\mbf{E}\left[f_2(R^l(t)){\phi\big(\mu(Y(t))+\sigma(Y(t))\epsilon(t)\big)\over\det(\sigma^{-1}(Y(t)))\phi\left(\epsilon(t)\right)}\Big|\sigma(\mcal{F}^{lR}_{t-1},\mcal{F}_t^Y)\right]}{F}{lR}{t-1}.\nonumber
\end{eqnarray}
A similar argument about independence of $\epsilon(t)$ and
$\mcal{F}^{lR}_{t-1}$ gives
\begin{eqnarray}
\lefteqn{\condE{\lambda_t^{-1}f_1(Y(t))f_2(R^l(t))}{F}{lR}{t-1}=}\nonumber\\
&&=\condE{f_1(Y(t))p_1\int_{\mbb{R}^N}f_2(\mu(Y(t))+\sigma(Y(t))\hat{\epsilon}){\phi\big(\mu(Y(t))+\sigma(Y(t))\hat{\epsilon}\big)\over\det(\sigma^{-1}(Y(t)))}\dd\hat{\epsilon}}{F}{lR}{t-1}\nonumber
\end{eqnarray}
and after the diffeomorphic change of variables
\begin{eqnarray}
&&=\condE{f_1(Y(t))p_1\mbf{E}\left[f_2(R^{l}(t))\big|\mcal{F}^{lR}_{t-1}\right]}{F}{lR}{t-1}\nonumber\\
&&=\condE{f_1(Y(t)){\psi\big(\tilde{y}-\alpha(\bar{y}-\tilde{y})+\sigma_Y\xi(t)\big)\over\det(\sigma^{-1}_Y)\psi(\xi(t))}\mbf{E}\left[f_2(R^{l}(t))\big|\mcal{F}^{lR}_{t-1}\right]}{F}{lR}{t-1}\nonumber\\
&&=\condE{\int_{\mbb{R}^K}f_1(\tilde{y}-\alpha(\bar{y}-\tilde{y})+\sigma_Y\hat{\xi}){\psi\big(\tilde{y}-\alpha(\bar{y}-\tilde{y})+\sigma_Y\hat{\xi}\big)\over\det(\sigma^{-1}_Y)}\dd\hat{\xi}\cdot\mbf{E}\left[f_2(R^{l}(t))\big|\mcal{F}^{lR}_{t-1}\right]}{F}{lR}{t-1}\nonumber\\
&&=\condE{\mbf{E}\left[f_1(Y(t))\big|\mcal{F}^{lR}_{t-1}\right]\cdot\mbf{E}\left[f_2(R^{l}(t))\big|\mcal{F}^{lR}_{t-1}\right]}{F}{lR}{t-1}\nonumber\\
&&=\mbf{E}\left[f_1(Y(t))\big|\mcal{F}^{lR}_{t-1}\right]\cdot\mbf{E}\left[f_2(R^{l}(t))\big|\mcal{F}^{lR}_{t-1}\right].\nonumber
\end{eqnarray}
Taking $f_1(z)=\mbf{1}_{A}(z)$ and $f_2(z)=\mbf{1}_{B}(z)$ for
$A,B\in\mcal{F}^{lR}_t$ we obtained
\begin{eqnarray}
\lefteqn{\mbf{P}\left(\{R^{l}(t)\in A\}\cap \left\{Y(t)\in B\right\}\big|\mcal{F}^{lR}_{t-1}\right)=\condE{f_1(R^{l}(t))f_2(Y(t))}{F}{lR}{t}=}\nonumber\\
&&=\condE{f_1(R^{l}(t))}{F}{lR}{t}\condE{f_2(Y(t))}{F}{lR}{t}=\mbf{P}\left(R^{l}(t)\in
A\big|\mcal{F}^{lR}_{t-1}\right)\mbf{P}\left(Y(t)\in
B\big|\mcal{F}^{lR}_{t-1}\right).\nonumber
\end{eqnarray}
\koniec

\dowod[Theorem \ref{tw:recursive_varrho}]  For an arbitrary Borel function $f$ let us take
\begin{eqnarray}
\lefteqn{\int_{\mbb{R}^K}f(z)\varrho_t(z)\dd
z=\barcondE{\Lambda_tf(Y(t))}{F}{lR}{t}=}\nonumber\\
&&=\barcondE{\Lambda_{t-1}{\phi\Big(\sigma^{-1}(Y(t))[R^{(l)}(t)-\mu(Y(t))]\Big)
\over\det(\sigma(Y(t)))\phi(R^{(l)}(t))}{\psi\Big(\sigma_Y^{-1}[Y(t)-\tilde{y}-\alpha(\bar{y}-\tilde{y})]\Big)\over\det(\sigma_Y)\psi(Y(t))}f(Y(t))}{F}{lR}{t}\nonumber\\
&&=\bar{\mbf{E}}\left[\Lambda_{t-1}\mbf{E}\left[{\phi\Big(\sigma^{-1}(Y(t))[R^{(l)}(t)-\mu(Y(t))]\Big)
\over\det(\sigma(Y(t)))\phi(R^{(l)}(t))}\right.\right.\nonumber\\
&&\qquad\qquad\qquad\qquad\left.\left.{\psi\Big(\sigma_Y^{-1}[Y(t)-\tilde{y}-\alpha(\bar{y}-\tilde{y})]\Big)\over\det(\sigma_Y)\psi(Y(t))}f(Y(t))\Bigg|\sigma(\mcal{F}^{lR}_t,\mcal{F}^Y_{t-1})\right]\Bigg|\mcal{F}_t^{lR}\right]\nonumber\\
&&=\barcondE{\Lambda_{t-1}\int_{\mbb{R}^K}{\phi\Big(\sigma^{-1}(z)[R^{(l)}(t)-\mu(z)]\Big)
\over\det(\sigma(z))\phi(R^{(l)}(t))}{\psi\Big(\sigma_Y^{-1}[z-\tilde{y}-\alpha(\bar{y}-\tilde{y})]\Big)\over\det(\sigma_Y)\psi(z)}f(z)\psi(z)\dd
y }{F}{lR}{t}.\nonumber
\end{eqnarray}
The inner process is the function of $Y(t-1)$ and $\Lambda_{t-1}$ so
we can integrate with respect to $\varrho_{t-1}(z)$ in the
following
\begin{eqnarray}
\int_{\mbb{R}^K}f(z)\varrho_t(z)\dd
z&=&\int_{\mbb{R}^K}\left(\int_{\mbb{R}^K}{\Phi(z,R^{l}(t))\over\phi(R^{l}(t))}\Psi(z,y)f(z)\dd
z\right)\varrho_{t-1}(y)\dd y\nonumber\\
&=&\int_{\mbb{R}^K}f(z){\Phi(z,R^{l}(t))\over\phi(R^{l}(t))}\left(\int_{\mbb{R}^K}\Psi(z,y)\varrho_{t-1}(y)\dd
y\right)\dd z.\nonumber
\end{eqnarray}
Since $f$ is the arbitrary function the identity
\ref{eq:recursive_filter} holds. \koniec
\section{The constant $v_N$ in the error estimate in lemma \ref{tw:main_quantiz_result}}
Changing the limits of integration in the integral $$\int_{[M_0,+\infty)^N}|x|^{-p}\dd x$$ to the polar ones
with
$$F(w,\alpha_1,\dots,\alpha_N)=\left[\begin{matrix}\hfill w\cos\alpha_1\cos\alpha_2\cos\alpha_3\dots\cos\alpha_{N-1}\cr\hfill w\sin\alpha_1\cos\alpha_2\cos\alpha_3\dots\cos\alpha_{N-1}\cr\hfill w\sin\alpha_2\cos\alpha_3\dots\cos\alpha_{N-1}\cr\vdots\cr\hfill w\sin\alpha_{N-2}\cos\alpha_{N-1}\cr\hfill w\sin\alpha_{N-1}\end{matrix}\right]$$
we obtain
\begin{eqnarray}
\lefteqn{\int_{[M_0,+\infty)^N}|x|^{-p}\dd
x=\int_{[M_0,+\infty)\times[0,2\pi]\times[-{\pi\over 2}{\pi\over
2}]^{N-2}}r^{-p}|\det F'|\dd
w\dd\alpha_1\alpha_2\dots\alpha_{N-1}=}\nonumber\\
&&\int_{[M_0,+\infty)}w^{-p}w^{N-1}\int_{[0,2\pi]}1\dd
\alpha_1\int_{[-{\pi\over 2}{\pi\over
2}]}\cos\alpha_2\dd\alpha_2\dots\int_{[-{\pi\over 2}{\pi\over
2}]}(\cos\alpha_{N-1})^{N-2}\dd\alpha_{N-1}.\nonumber
\end{eqnarray}
Let us denote $I_n\colon=\int_{-{\pi\over 2}}^{\pi\over 2}\cos^nt\dd t.$
Note that $I_0=\pi$ and $I_1=2$. Integrating by parts
\begin{eqnarray}
\int_{-{\pi\over 2}}^{\pi\over 2}\cos t\cos^{n-1}t\dd t&=&\sin t\cos^{n-1}t\Big|_{-{\pi\over 2}}^{\pi\over 2}+\int_{-{\pi\over 2}}^{-{\pi\over 2}}\sin^2t\cos^{n-2}t\dd t\nonumber\\
I_n&=&0+nI_{n-1}-nI_n,
\end{eqnarray}
thus
\begin{eqnarray}
n=2k&\Rightarrow&I_n={n!!\over(n+1)!!}\pi,\nonumber\\
n=2k+1&\Rightarrow&I_n=2{n!!\over(n+1)!!}.\nonumber
\end{eqnarray}
Then
\begin{eqnarray}\nonumber
v_N&=&{2\pi\over N-p}\prod_{j=2}^{N-1}\int_{[-{\pi\over 2},{\pi\over 2}]}(\cos t)^{j-1}\dd t\nonumber\\
&=&{2\pi\over N-p}\prod_{j=1}^{N-2}2^{j\mathrm{mod}2}\pi^{(j+1)\mathrm{mod}2}{j!!\over(j+1)!!}\nonumber\\
&=&\left\{\begin{matrix}{2\pi\over (N-p)(N-2)!!}(2\pi)^{{N-2\over
2}},\hfill&\mathrm{if}\ N=2k\hfill\cr{2\pi\over
(N-p)(N-2)!!}2^{{N-3\over 2}}\pi^{{N-3\over
2}+1},\hfill&\mathrm{if}\ N=2k+1\hfill\cr\end{matrix}\right.
\end{eqnarray}
\section*{Acknowledgements} 
The paper is an extension of the part of my doctoral thesis completed under supervision of Prof \L ukasz Stetter. I would like to thank the participants of the Third General AMAMEF Conference ``Advances in Mathematical Finance'', May 5-10, 2008, Pitesti, Romania, for their valuable comments to the preliminary results which were presented there. I also expresses my gratefulness to Tom Hurd and the Fields Institute in Toronto for invitation to the postdoctoral program and the financial support that contributed a lot to essentially improving and finishing the paper. Many thanks to my wife Ania for the interminable support in my research.
\bibliographystyle{elsarticle-num}

\end{document}